\documentclass[preprint,showpacs,preprintnumbers,amsmath,amssymb]{revtex4}
\usepackage{graphicx}
\usepackage{bm}
\usepackage{color}
\usepackage{amssymb}
\usepackage{epsfig}

\begin{document}
\topskip 20mm
\title{Criticality in Brownian ensembles}
\author{ Suchetana Sadhukhan and Pragya Shukla}
\affiliation{Department of Physics, Indian Institute of Technology,
Kharagpur, India}
\date{\today}
\begin{abstract}	
	

A Brownian ensemble appears as a non-equilibrium state of transition from one universality class of random matrix ensembles to another one.   The parameter governing the transition  is in general size-dependent, resulting in a rapid approach of the  statistics, in infinite size limit,  to one of the two universality classes.  Our detailed analysis however reveals  appearance of a  new scale-invariant spectral statistics, non-stationary along the spectrum, associated with multifractal eigenstates and different from the two end-points  if the transition parameter becomes  size-independent. The number of such critical points  during  transition is governed by a competition between  the average perturbation strength and the local spectral density. The results obtained here have  applications to wide-ranging complex systems e.g. those modeled by multi-parametric Gaussian ensembles or column constrained ensembles.  




\end{abstract}


\maketitle

\section{Introduction}

Recent studies of  the localization to delocalization transitions e.g many body localization, Anderson localization  and random graphs indicate a common mathematical structure underlying the statistical fluctuations of their linear operators \cite{krav, psrp, bgg}. The structure belongs to that of a Rosenzweig-Porter (RP) ensemble \cite{rp}, or, equivalently, to a specific type of  Brownian ensemble (BE) i.e  the one intermediate between Poisson and Gaussian ensembles \cite{psand}.  This indicates a crucial but so far hidden statistical connection of the BEs with systems undergoing localization-delocalization transition. It is therefore natural to search for the criticality in BEs which motivates the present study.

A Brownian ensemble in general refers to an intermediate state of perturbation of a stationary random matrix ensemble by another one of a different universality class \cite{dy, me, ap}.  The type of a BE, appearing during the cross-over, depends on the nature of the stationary ensembles and their different pairs may give rise to different BEs \cite{ap,  psijmp}. Similar non-stationary states may also arise in other matrix spaces e.g. unitary matrix space e.g. due to a perturbation of a stationary circular ensemble by another one \cite{apps, sp, vp, psnher}.  The BEs  have been  focus of many studies in past decades  (for example see \cite{sp, vp, pslg} and the references therein) and a great deal of analytical/ numerical information is already available about them. 
%
%
However very few of these studies \cite{alt, psand, krav} probed the critical aspects of the BEs which refers to a behavior different from the two stationary limits in infinite matrix size limit \cite{mj}. The search of criticality in BEs is important for several reasons. For example,  the analytical study \cite{ps-all} indicate that the statistical fluctuations of a wide range of complex systems are analogous to that of a Brownian ensemble, subjected to similar global-constraints,  if their complexity parameters are equal irrespective of other system-details. 
(The complexity parameter is  a function of the distribution parameters of the ensemble or alternatively a function of the average accuracy of the matrix elements, measured in units of the mean-level spacing). A recent study \cite{ss1} also reveals the connection of the BE to the random matrix ensembles with column/row constraints; the latter appear in diverse areas e.g bosonic Hamiltonians such as  phonons, and spin-waves in Heisenberg and XY ferromagnets, antiferromagnets, and spin-glasses,  euclidean random matrices, random reactance networks, financial systems and Internet related Google matrix  etc. The knowledge of criticality in BEs can therefore be helpful in its search in other related ensembles.

The criteria for the critical statistics of energy levels and eigenfunctions was first introduced to an ensemble of disordered Hamiltonians undergoing  localization to delocalization transition \cite{shk}. It has long been believed that a fractional value of the spectral compressibility and multifractal behavior of the eigenfunctions are  signatures of the  criticality in the ensemble \cite{ckl, azks}. In fact these measures were used to claim the analogy of the Anderson ensemble (AE) at metal-insulator transition with that of the Power law random banded matrix (PRBM) ensemble \cite{prbm}. The study \cite{psand, pssymp} indicates that  the statistics of both of these ensembles can be mapped to that of the BE$_{p \to o}$ (with subscript indicating the two end points i.e Poisson and Gaussian orthogonal ensemble); the BE is therefore expected to show similar critical features too. 
This is however at variance with  another study \cite{krav} which suggests the criticality in RP ensemble (and therefore in BE$_{p \to o}$) is different from AE and PRBME; this suggestion is based on a perturbative analysis  of the eigenfunction fluctuations and two point spectral correlation (also see \cite{fkpt, alt, fgm, ls, ks} for related studies).
The need for a clear answer motivates us to pursue an analytical calculation of the spectral compressibility and multifractality for the BEs.  Although our approach is applicable for a generic BE  of both Gaussian or Wishart type (i.e intermediate between an arbitrary initial condition and the  Gaussian/ Wishart type stationary ensembles,  these measures so far seem to be relevant in context of the ensembles undergoing localization to delocalization transition. To strengthen and support the theoretical analysis,  we probe the  behavior by numerical route too but that is confined to  the Gaussian  BEs  between Poisson to  GOE only.

The paper is organized as follows. Section II briefly introduces the Brownian ensembles in Hermitian matrix spaces.  
The diffusive dynamics for their eigenvalues and the eigenfunction components was analyzed in detail in \cite{sp}, \cite{pswf} and \cite{pslg}, respectively. This information is used in sections III and IV to derive the parametric dependence of the criticality measures i.e spectral compressibility, the multifractality spectrum and eigenfunction correlations at two different energies. Here we also discuss the conditions under which they become critical. Although the results of section III and section IV are applicable for arbitrary initial conditions, the  main interest in these measures arises, so far, from the quest to characterize the localization to delocalization transition. This motivates us to focus on the corresponding BE i.e  BE$_{p\to o}$ in subsequent sections and numerically verify our theoretical results for  them. Section V very briefly reviews the basic formulation for these BEs and presents the details of our numerical analysis.  Section VI analyzes the reasons for  the  seemingly contradictory claims of the studies \cite{psand} and \cite{krav}. We believe it can be explained on the basis of a rate of change of  the local density of states which affects the local statistical fluctuations.   Section VII concludes with summary of our main results and open questions.

\section{Brownian ensembles: the definition}

 Introduced by Dyson to model the statistical behavior of systems with partially broken symmetries and/or approximate conservation laws \cite{dy, me}, a BE was originally based on the assumption of Brownian dynamics of matrix elements due to thermal noise.  But currently a BE is also  described as a single parameter governed diffusive state of the matrix elements of a randomly perturbed  stationary ensemble \cite{me, ap, sp, psijmp}.  Consider an ensemble of $N_a \times N$  rectangular matrices $A(\lambda)=\sqrt{f} (A_0+\lambda V)$ with $f=(1+\lambda^2)^{-1}$ \cite{sp, pslg} and  matrices $A_0$ and $V$ distributed with probability densities $\rho_0(A_0)$ and $\rho_v(V)$. 
As clear, $A=A_0$ for $\lambda \rightarrow 0$, $A \rightarrow V$ for $\lambda \rightarrow \infty$. 
The ensemble of rectangular matrices $A$ can lead to three important classes of $N \times N$ Hermitian matrix ensembles (i) Gaussian ensembles of matrices $H =A+A^{\dagger}$ with $N=N_a$, (ii) Wishart ensembles  with matrices $L=A^{\dagger} A$ (also referred as Laguerre ensembles), and, (iii) Jacobi ensembles of matrices $S$ which approach a form $S= (A^{\dagger} A+ B^{\dagger}B)^{-1/2} \; (B^{\dagger} B- A^{\dagger}A) \; (A^{\dagger} A+ B^{\dagger}B)^{-1/2}$. Our theoretical analysis in this paper is confined only to the first two ensembles.


A variation of strength $\lambda$ of the random perturbation $V$ leads to diffusion of the matrix elements $A_{kl}=\sqrt{f} (A_{0;kl}+\lambda V_{kl})$  which, by a suitable choice of $\rho_v(V)$, can be confined to a finite space.  For example, for the Gaussian density of the $V$-ensemble, the  Markovian character of the dynamics is preserved if considered in terms of a rescaled parameter $Y$ given by the relation $f={\rm e}^{-2Y}$ \cite{sp}. 
For $\rho_v(V) =\left(\frac{1}{2 \pi v^2}\right)^{\beta N_a N/2} {\rm e}^{-{1\over 2v^2} \; {\rm Tr} (V V^{\dagger})}$,  the diffusion equation for the matrix elements of $X$ (with $X \equiv H$ or $L$) can explicitly be derived \cite{sp, pslg} (with $\beta=1,2$ for $X$ as real-symmetric or complex Hermitian, respectively). As discussed in \cite{sp, pswf, pslg}, this in turn leads to the $Y$-governed diffusion equation  for the JPDF (joint probability distribution function) of their $N$ eigenvalues $e_k$, $k=1 \to N$ and corresponding eigenfunctions. A direct integration of the JPDF diffusion equation over all eigenfunctions and $N-n$ eigenvalues leads to the diffusion equation for the $n^{th}$ order level-density correlation $R_n(e_1, e_2, \ldots, e_n)$. The measure $R_1(e)$ is also referred as the ensemble average level density, with its fluctuations described by $R_n$, $n >1$. 
As discussed in \cite{ap}, the crossover in $R_1$  occur at a  scale $Y \sim N \Delta_e^2$ with $\Delta_e$ as the local mean level spacing. The crossover in $R_n$ is however rapid and occurs at scale $Y \sim \Delta_e^2$. 
For comparison of  local spectral fluctuations around the level density, therefore, a rescaling of the eigenvalues by local mean level spacing $\Delta_e(e)=R_1^{-1} $ (also referred as unfolding) is necessary. This however leads to a rescaling of both $R_n$ as well as the  crossover parameter $Y$, with new parameter $\Lambda_e$ given as 
\begin{eqnarray}
\Lambda_e(Y,e)={e^{\nu} \; (Y-Y_0) \over  \Delta_e^2}
\label{alm1}
\end{eqnarray}
with $\nu=0, 1$ for Gaussian and Wishart ensembles respectively and $Y_0$ is value of $Y$ for initial ensemble $A=A_0$.

As discussed in \cite{ps-all, psand}, $\Lambda_e$ also appears as the single parameter governing the spectral statistics of a multi-parametric Gaussian ensemble (which includes  Gaussian BEs as a special case); $Y$ in this case is the function of all  ensemble parameters, thus containing information about the ensemble complexity. $\Lambda_e$ is therefore also referred as the spectral complexity parameter.  

%
%




It must be emphasized here that, before unfolding, the correlations in a BE depends on two parameters, namely, local mean level density and perturbation parameter $Y$.  Although the unfolding maps the local mean level density to a constant, it however introduces a spectral-scale dependence in the rescaled evolution parameter $\Lambda_e$. The evolution of $R_n$ for $n >1$ is therefore different 
at different spectral scales which implies the non-stationarity of local fluctuations of the BE. 
This is different from the stationary ensembles in which correlations $R_n$ depends only  on 
one parameter i.e local mean level density; the unfolding in this case results in a constant local level density and as a consequence, $R_n$ become independent of spectral scale. 





Contrary to spectral correlations, the local eigenfunction correlations in a BE are  governed by different rescaling  of $Y$ sensitive to the measure under consideration \cite{pswf, pslg}.  This results in varying cross-over speeds for the eigenfunction fluctuations and is in fact an indicator of the multiple scale dependence of the local eigenfunction intensity.








\section{Signatures of criticality in spectral statistics}


In general, the criticality in a joint probability distribution function (JPDF) of the eigenvalues can be defined as follows.  A one-parameter scaling behavior of the distribution $P(\{e\})$ implies the existence of a universal distribution $P^*(\{e\})= {\rm lim}_{N\rightarrow \infty} \; P(\{e\},\Lambda_e)$ if the limit $\Lambda^*={\rm lim}_{N\rightarrow \infty}\;  \Lambda_e (N)$ exists \cite{mj}.    Thus the size-dependence of $\Lambda_e$ plays a crucial role in locating the critical point of statistics. Let $|Y-Y_0| \propto N^{\alpha}$ and $\Delta_e \propto N^{\eta} $, eq.(\ref{alm1}) then  gives  $\Lambda_e \propto N^{\alpha-2\eta}$. A variation of size $N$ in finite systems then leads to a smooth crossover of spectral statistics  between an initial state ($\Lambda_e \rightarrow 0$) and the equilibrium ($\Lambda_e \rightarrow \infty$); the intermediate statistics belongs to an infinite family of ensembles, parameterized by $\Lambda_e$. However, for system-conditions leading to $\alpha=2\eta$, the spectral statistics becomes universal for all sizes, $\Lambda_e$ being $N$-independent; the corresponding system conditions can then be referred as the critical conditions (or point).   It should be stressed that the  system conditions satisfying the critical criteria  may not exist in all systems; the critical statistics therefore need not  be a generic  feature of all systems.

At critical value $\Lambda^*$, $R_n(r_1)$ (for $n >1$) and therefore all  spectral fluctuation measures are different from the two end points of the transition i.e $\Lambda_e=0$ and $\infty$. Any of them can therefore be used, in principle, as a criteria for the critical statistics. 
A direct theoretical or numerical study of the JPDF of eigenvalues or the correlations $R_n$  is however  not the most suitable approach for the analysis. This has in past  led to introduction of many alternative measures \cite{mj} e.g. nearest neighbor spacing distribution, number variance, spectral rigidity etc. \cite{me}. An important aspect  of these measures is their spectral scale dependence.     As mentioned in previous section, the spectral correlations in BEs retain their energy-dependence through $\Lambda_e$ even after unfolding and are non-stationary  i.e vary along the spectrum \cite{gmrr}. Any criteria for the  criticality in the spectral statistics can then be defined only  locally i.e within the energy range, say $\delta e_c$, in which $\Lambda_e$ is almost constant. From eq.(\ref{alm1}), 
${{\rm d}\Lambda_e\over {\rm d}e} = 2 (Y-Y_0) R_1 {{\rm d}R_1\over {\rm d}e}$ which implies that $\delta e_c$ is large only for regions where $R_1(e) \gg 2 {{\rm d}R_1\over {\rm d}e} $. 



The $\Lambda_e$- governed diffusion of the eigenvalues  subjects the local spectral fluctuation measures  also to undergo a similar dynamics. To determine their behavior at the critical point,  it is necessary to first obtain the evolution equations for the relevant measures. The spectral compressibility being a popular measure as well as related to  other criteria for spectral criticality,  here we consider its evolution.

\subsubsection{Spectral compressibility and its evolution}


As mentioned in section I,  the spectral compressibility $\chi$ is an often used criteria for the criticality statistics in the ensembles of disordered Hamiltonians. A characteristic of the  long-range correlations of levels, it is defined as,  in a range $r$ around energy $e$, 
\begin{eqnarray}
\chi(e, r) =1- \int_{-r}^{r} (1-R_2(e, s) ) \; {\rm d}s.
\label{chi}
\end{eqnarray}
where $R_2(e,r) \equiv R_2(e,e+r)$ is the two point level density correlation at an energy $e$. As $R_2(e,r)$ is related to another $2$-point measure, namely, the number variance $\Sigma_2(e,r)$, (the variance in the number of levels in an interval of $r$ mean number of levels),  $\chi$  can also be expressed as the $r$-rate of change of $\Sigma_2(e,r)$
\cite{mj, azks, ckl}): $\Sigma_2(r) \sim \chi r$ for large $r$ with $0 <\chi < 1$.
(As the  interest is often in  large $r$-behavior of $\chi$ at a fixed energy $e$, its dependence on energy $e$ is usually suppressed). In \cite{ckl}, $\chi$ was suggested  to be  related to the multifractality of  eigenfunctions: $\chi= \frac{d-D_2}{2 d}$ with $D_2$ as the fractal dimension and $d$ as the system-dimension.   However numerical studies indicated the result to be valid only in the weak-multifractality limit \cite{evers}. 
Later on, another criteria was introduced in terms of the level-repulsion (an indicator of short range correlation), measured by nearest- neighbor spacing distribution.  The study \cite{shk} showed that the nearest-neighbor spacing distribution $P(s)$ turns out to be a universal hybrid of the GOE at small-$s$ and Poisson at large-$s$, with an exponentially decaying tail: $P(s) \sim {\rm e}^{-\kappa s}$ for $s \gg 1$. Here $\kappa$ is a constant and is believed to be related to $\chi$: $\kappa=\frac{1}{2 \chi}$.

For the spectrum of uncorrelated levels (no level repulsion) i.e Poisson ensemble,  $R_2(e, r)=1$ which  gives $\chi=1$.  But for a classical ensemble (e.g. Gaussian orthogonal or Unitary ensembles), 
the well-known sum rule $\int_{-N/2}^{N/2} (1-R_2(e, r)) {\rm d}r=1$    gives $\lim_{r \rightarrow N/2} \chi(e, r)=0$;  this implies that  a classical ensemble corresponds to the maximum level repulsion (i.e zero compressibility)  in the related symmetry class \cite{me, psijmp}.  
Clearly if $\lim_{r \rightarrow N/2} \;  \chi(e, r) \not= 0, 1$, it characterizes a spectrum different from classical ensembles as well as uncorrelated spectrum. This characterization however is suitable only for the stationary spectrum (where unfolded spectral correlations are independent of the location along the energy axis).  In case of the non-stationarity, the statistics varies along the energy-axis (even after unfolding)  and one can at best define a local compressibility within an energy range $E_{st}$ ($\ll$ total spectrum width) in which the local stationarity is valid. 
%
This led to introduction of the following criteria for criticality: the spectral statistics is believed to be critical if 
\begin{eqnarray}
\lim_{r \rightarrow \infty} \;  \lim_{N \rightarrow \infty} \; \chi(e, r) \not=0, \not=1.
\label{chi-c}
\end{eqnarray} 
(Note the order of limits on $r$ and $N$ are non-interchangeable.
This leads to technical issues in numerical search for criticality in $\chi$: the total number of levels $N$ in the spectrum
 being finite,  the maximum range of allowed $r$ is $r \le N_{st} \ll N$, with $N_{st}={E_{st}\over \Delta_e}$ and it is not easy to realize a large $r$ limit). 
To determine  $\chi(e, r)$   from eq.(\ref{chi}),  a prior information of $R_2$ is needed. Unfortunately  an exact form of $R_2$  is known for very few BE cases e.g Poisson to GUE, GOE to GUE, uniform to GUE \cite{ap}. But  the condition for a fractional value of $\chi$ can be obtained by general considerations. 
As discussed in \cite{ap,sp},  a variation of perturbation strength of the BE subjects $R_2(r)$ to undergo diffusion, described as 
\begin{eqnarray}
{\partial R_2 \over\partial \Lambda_e} &=& 
2 {\partial \over \partial r} \; \left[{\partial R_2\over \partial r}
- \beta {R_2 \over r} -\beta \; 
\int_{-N/2}^{N/2} {R_3(0,x, r) \over x} \; {\rm d}x \; \right].
\label{r2}
\end{eqnarray}
with $R_3(0,x,r)$ as the $3$-point level-density correlation and $\Lambda_e$ given by eq.(\ref{alm1}). Note the above equation is applicable only locally i.e within spectral scale in which $R_1(e)$ is almost constant and $R_2$ is translationally invariant. The latter allows one to write $R_2(e,r) = R_2(r)$ but $e$-dependence enters through $\Lambda_e$. 
%
%
By differentiating eq.(\ref{chi}) with respect to $\Lambda_e$, followed by a substitution of eq.(\ref{r2}) and subsequent repeated partial integrations, leads to following approximated closed form equation for $\chi(r)$ (suppressing $e$-dependence of $\chi$ for clarity of presentation)
\begin{eqnarray}
{\partial \chi \over\partial \Lambda_e} &=& - 4 \; \left( \frac{\beta}{r} - {\partial\over \partial r}\right)R_2(r; \Lambda_e)  - 4 \; \beta \;   
\int_{-N/2}^{N/2} {R_3(0,x,r; \Lambda_e) \over x} \; {\rm d}x 
\label{chi1}
\end{eqnarray}

An integration over $\Lambda_e$ of the above equation now gives 
\begin{eqnarray}
\chi  (r; \Lambda_e) 
&=& \chi(r; 0) - \left[4 \; \left(\frac{\beta}{r} - {\partial\over \partial r}\right) \; \phi_1(r; \Lambda_e) \right] -  
4 \; \beta \;   \phi_2(\Lambda_e)  
\label{chi2}
\end{eqnarray}
where
$\phi_1(r; \Lambda_e) = \int_{0}^{\Lambda_e} \; {\rm d} t \; R_2(r, t), \label{u}$ and 
$\phi_2(\Lambda_e)  = \int_{0}^{\Lambda_e} \; {\rm d} t \int_{-N/2}^{N/2} \; {\rm d}x \; {R_3(0,x; r, t) \over x}$.
Further simplification of eq.(\ref{chi2}) is possible based on following points 
(i) $R_3$ can also be expressed in terms of $R_2$:
$R_3(0,x,r) = Y_3(0,x,r) + R_2(x) + R_2(r) + R_2(r-x) -2$ with $Y_3(0,r,x)$ as the $3^{rd}$ order cluster function \cite{me, ap}. 
(ii) the range of integral over $x$ in the definition of $\phi_2$ varies from $-N/2$ to $N/2$ and our interest is in the limit $N \to \infty$ followed by $r \to \infty$, 
(iii) as $R_3$ varies from $0 \to 1$, the main contribution to the integral over $x$ in $\phi_2$ comes from the neighborhood of $x=0$. Thus although the range of integration 
$x$ varies from $-N/2$ to $N/2$, one needs to concern only with small $x$-values, 
(iv) the cluster function $Y_3(0,r,x)$ vanishes if $x$ or $r$ or $|x-r|$ becomes large in comparison to the local mean level spacing \cite{me}. 
 In large $r$-limit,  therefore,  one can approximate 
$R_2(r) \approx R_2(r-x) \to 1$ which leads to $\lim_{r \to \infty} R_3(0,x,r, t) \approx  R_2(x, t)$. Using the latter, $\phi_2$ can be expressed in terms of $\phi_1$:
 $\phi_2(\Lambda_e) =  \int_{-N/2}^{N/2} \; {\rm d}x \; {\phi_1(x, \Lambda_e) \over x}  $. 
The lack of energy level correlations at large $r$ i.e $R_2(r,t) \to 1$ for arbitrary $t$, also gives 
\begin{eqnarray} 
\lim_{r \to \infty, N \to \infty} \phi_1(r; \Lambda_e) =\int_{0}^{\Lambda^*} {\rm d} t \; \left(\lim_{r \to \infty} R_2(r, t) \right) \approx \Lambda^*.
\label{phip}
\end{eqnarray}
  In the ordered limit $r \to \infty, N \to \infty$, eq.(\ref{chi2}) can now be reduced to following form
 \begin{eqnarray}
\lim_{r \to \infty} \chi  (r; \Lambda^*)  =
\lim_{r \to \infty, N \to \infty} \chi  (r; \Lambda_e) 
&=& \lim_{r \to \infty, N \to \infty}  \chi(r; 0) - 4 \; \beta \; {\mathcal I}_0 
\label{chi3}
\end{eqnarray}
with $\Lambda^*= \lim_{N \to \infty} {\Lambda}_e$ and  ${\mathcal I}_0  = \lim_{r \to \infty, N \to \infty} \; \phi_2(\Lambda_e) = \int_{-\infty}^{\infty} \; {\rm d}x \; {\phi_1(x, \Lambda^*) \over x}  $. 
Further insight however can be gained by the following reasoning. As 
${\mathcal I}_0 =  \int_{0}^{\Lambda^*} \; {\rm d} t \; \int_{-\infty}^{\infty} \; {\rm d}r \;  {R_2(r, t) \over r}  $, the dominant contribution to the integral over $r$ comes from the region near $r=0$. 
(This can also be seen as follows. In general, the eigenvalues at distances more than a system-specific spectral-range, say $E_c$ around $e$,  are uncorrelated. Here  $E_c$ is a crucial spectral-range, hereafter, referred as the Thouless energy, as in the context of disordered systems in which case usually $E_c \sim \Delta_e$. This  implies $R_2(r,t) \to 1$ for $r > N_e$ where $N_e=E_c/ \Delta_e$ , one can write $\int_{-\infty}^{\infty} \; {\rm d}r \;  {R_2(r, t) \over r} = \int_{-\infty}^{-N_e}  {{\rm d}r \over r} + \int_{N_e}^{\infty}  {{\rm d}r \over r} + \int_{-N_e}^{N_e} \; {\rm d}r \;  {R_2(r, t) \over r}  $. Due to symmetry, the first two terms  cancel out leaving only the last term.)  
Thus ${\mathcal I}_0$ is sensitive to the short range behavior of $R_2$ i.e degree of level-repulsion in the spectrum. 

It is worth noting here the advantage of eq.(\ref{chi3}) over eq.(\ref{chi}):  although calculation of $\chi$ by both eq.(\ref{chi}) and eq.(\ref{chi3}) depends on a prior knowledge of $R_2$ but later requires only its small-range behavior which can easily be derived from eq.(\ref{r2}), for arbitrary initial conditions, by  neglecting the integral term. As an example consider the BE intermediate to Poisson and Gaussian orthogonal ensemble (GOE); the small-r solution of eq.(\ref{r2}) for this case can be given as $R_2(r, \Lambda) \approx \left({\pi \over 8 \Lambda}\right)^{1/2} \; r \; {\rm e}^{-r^2/16 \Lambda} \; I_0\left({r^2 \over 16 \Lambda}\right)$ where $I_0$ is the modified Bessel function. Substitution of the latter in ${\mathcal I}_0$, leads to
\begin{eqnarray} 
\chi \approx 1 - 4  \; \sqrt{2 \pi} \; \eta_0 \; \Lambda^*
\label{cth}
\end{eqnarray}
where $\eta_0= \int_{-N_e}^{N_e} {\rm e}^{-r^2} \; I_0\left({r^2}\right) \; {\rm d}r \approx \sqrt{\pi} $ with $\chi(r,0) =1$ in Poisson limit.

Further insight in the large-$r$ behavior of $\chi(r; \Lambda_e)$ can be derived 
through a $\Lambda_e$ governed evolution equation in the spectral-region. The 
steps are as follows. 
Eq.(\ref{chi}) gives,$1+ {1\over 2} \frac {\partial \chi(r)}{\partial r}=R_2(r) $.
In large-$r$ limit, this leads to the  approximation 
\begin{eqnarray}
\int_{-\infty}^{\infty} {R_3(0,x,r; \Lambda_e) \over x} \; {\rm d}x \approx 
\int_{-\infty}^{\infty} {R_2(x,\Lambda_e) \over x} \; {\rm d}x 
= \int_{-\infty}^{\infty}  {1\over 2 x} \frac {\partial \chi(x)}{\partial x} \; 
{\rm d}x.
\label{r3c}
\end{eqnarray} 
%
Substitution of above relations in eq.(\ref{chi1}) gives $\Lambda_e$ governed evolution of $\chi(r)$ for large $r$ (with $\chi(\pm \infty) $ as constants):
\begin{eqnarray}
{\partial \chi \over\partial \Lambda_e} &=& \frac{-4 \beta}{r} - \frac{2 \beta}{r} \;  
{\partial \chi\over \partial r} + 2  \frac {\partial^2 \chi}{\partial r^2} 
- 2 \; \beta \; \int_{-\infty}^{\infty} {\chi(x) \over x^2} \; {\rm d}x 
\label{chi4}
\end{eqnarray}
As $0 < \chi(r; \Lambda_e) \le 1$,  the $1st$ and $2nd$ term on the right side of the above equation can be neglected for large $r$ and its integration over $\Lambda_e$ gives 
$\lim_{r \rightarrow \infty} \chi  (r; \Lambda_e) = \lim_{r \rightarrow \infty} \chi  (r; 0) - 2 \; \beta  \; \phi_3(\Lambda_e) $
with $\phi_3(\Lambda_e)=\int_0^{\Lambda_e} {\rm d}t \; \int_{-\infty}^{\infty} \; {\rm d}x \; {\chi(x; t) \over x^2}  $ 
(assuming $\frac {\partial^2 \chi}{\partial r^2}  \ll 1$ for large $r$). This reveals a bootstrapping tendency of $\chi(r)$ i.e the dependence of $\chi$ at large $r$ on its behavior near small $r$. Also note as both $\Lambda_e$ and $\Lambda^*$ are dependent on spectral scale $e$, $\chi$ is in general non-stationary along the spectrum.

\section{Signatures of criticality in eigenfunction statistics}


The basis-variant nature of an ensemble, which is often the case at the critical point,  implies a correlation between the eigenvalues and the eigenfunctions. The special features of the spectrum at the criticality are therefore expected to manifest in eigenfunctions too.  For example, as indicated by many studies of the localization $\to$ delocalization transitions, the eigenfunctions within spectral range supporting critical statistics have multifractal structure.
This has motivated three main criteria for the criticality in the eigenfunction fluctuations, namely, inverse participation ratio, multifractality spectrum and eigenfunction correlations at different energy.  Here we analyze these measures in context of the Brownian ensembles.

\subsubsection{Inverse participation ratio and its evolution}


The criticality  in the wavefunctions is believed to  manifest through large fluctuations of their amplitudes at all length scales and  is often characterized by an infinite set of critical exponents related to the scaling of the moments of the wave-function intensity $|\Psi(r)|^2$ with system size \cite{evers, mj}. The $q^{\rm th}$ moment $I_q$ of the wave-function intensity $|\Psi(r)|^2$, also known as $q^{\rm th}$ inverse participation ratio is defined as 
$I_q =\int {\rm d}r |\Psi(r)|^{2q}$ (equivalently $I_q =  \sum_n |\Psi_n|^{2q}$ in a $N$-dimensional basis  with $\Psi_n$ as the $n^{th}$ component of wavefunction $\Psi$). As revealed by the critical point studies of many disordered systems, an ensemble averaged $I_q $ reveals an anomalous scaling with size $N$:
$\langle I_q \rangle = N \; \langle \mid \Psi \mid^{2q}  \rangle \sim N^{-\tau_q/d}$
with $\langle . \rangle$ implying an ensemble average with $d$ as the system dimension; note $d=1$ for a BE. Here $\tau_q$ is a non-decreasing convex function with $ \tau_0=-d,\tau_1=0$.

The continuous set of exponents $\tau_q$ are related to the generalized
 fractal dimension $D_q$ of the wave-function structure: $\tau_q=(q-1)D_q$.
At critical point, $D_q$ is a non-trivial function of $q$, with  $D_q=d$ and $D_q=0$ for the eigenfunctions extended in a $d$-dimensional space and for completely localized ones, respectively. Further,
$\tau_q$ is also related to anomalous dimension $\Delta_q$ which distinguishes a multifractal state from an ergodic one and also determines the scale-dependence of the wave-function correlations: $\tau_q = d(q-1) + \Delta_q$ with  $ \Delta_0=\Delta_1=0$  \cite{evers}. 





For spectral regions with almost constant level density, the parametric-evolution of the average inverse participation ratio for a generic BE of Gaussian or Wishart type can be given as \cite{pslg, pswf}
\begin{eqnarray}
\overline{\langle  I_q(\Lambda_I)\rangle}= {\rm e}^{- t_2 \; \Lambda_I}\left[ \overline{\langle  I_q(0)\rangle} + 
t_1 \;  \int_0^{\Lambda_I}\overline{ \langle  I_{q-1}(r) \rangle} \; {\rm e}^{ t_2 \; r} {\rm d}r \right]
 \label{iu1}
\end{eqnarray}
with symbol $\overline{x}$ implying a local spectral averaging of a variable $x$. 
Here $t_1(q) = {2(q-1)+ \beta \over \beta \; } \langle|\Psi(r)|^2\rangle_e$,  $t_2(q) = 1+{1\over q \; {\mathcal K_2}} \left( \left({2\over \beta} \right)^{\nu} +{\nu N \over E_c} \right)$ and $\Lambda_I = q \; \beta \; {\mathcal K}_2 \;  (Y-Y_0)  $, ${\mathcal K}_s \approx {2^s \; N \over E_c^s} \; e^{\nu} $ and  $\nu=0, 1$ for the Brownian ensembles of Gaussian and Wishart type, respectively. 

The above equation clearly indicates the dependence of  $\overline{\langle  I_q(\Lambda_I)\rangle}$ on the spectral scale $e$ and system size $N$.  For finite but large $\Lambda_I$, it  can further be approximated as 
$\overline{ \langle  I_q(\Lambda_I)\rangle} \approx  \prod_{k=2}^q {t_1(k) \over t_2(k)} +O({\rm e}^{-t_2 \; \Lambda_I} )$.
With ${\mathcal K}_2 > {\mathcal K}_1 \gg 1$ (for large $N$), implying $t_2 \to 1$,  the above gives 
$\langle I_2 \rangle \approx {\beta +2\over \beta \; \xi}$ 
where $\xi$ is  the average localization length  in case of the localized eigenfunctions: $\xi \approx  {1\over \langle|\Psi(r)|^2 \rangle_e}$; this is in agreement with other studies \cite{ravin}. Further note, for $\Lambda_I \rightarrow \infty$,   $\overline{\langle  I_q \rangle}$ approaches a correct steady state limit, namely, XOE or XUE with X $\equiv$ L or G: 
 $\overline{\langle  I_q \rangle}  ={(2q)!\over 2^q q!} N^{1-q}$ for $\beta=1$ and 
$\overline{\langle  I_q \rangle} =q! N^{1-q}$ for $\beta=2$ \cite{evers}.   

As discussed in \cite{pslg}, the local intensity $\langle |\Psi(r)|^2 \rangle_e$ (given by $N^{-1} \; \langle u(r) \rangle$ in \cite{pslg}) depends on the perturbation strength $Y-Y_0$ of a BE and is different for Gaussian and Wishart ensembles. For later reference, here we mention the result for a Gaussian BE: 
 $\langle   |\Psi(r)|^2 \rangle_e \propto {1\over N \sqrt{Y-Y_0}}$. For a 
 BE appearing during Poisson to GOE or GUE, and, with  $Y-Y_0 \sim N^{-\gamma}$, this gives  $\Lambda_I \sim 
{N^{1-\gamma} \over E_c^2}$ and $\overline{\langle I_q \rangle} \sim  N^{(\gamma-2)(q-1)/2}$ for $q >0$. A comparison of the above result with $\overline{\langle  I_q(\Lambda_I)\rangle} \sim N^{-\tau_q}$ then gives, for $q >0$,
\begin{eqnarray}
\tau_q \approx {1\over 2} (q-1)(2-\gamma).
\label{tau1}
\end{eqnarray}
This in turn implies  all the fractal dimensions for large but finite $\Lambda_I$ of the BE are same: $D_q \approx {(2-\gamma)\over 2}$. 

{


\subsubsection{Diffusion of multifractality spectrum}

A well-known criteria for the multifractality  is the singularity spectrum $f(\alpha)$: it is defined as 
the fractal dimension of set of those points $r$ at which $|\psi(r)|^2 \sim N^{-\alpha/d}$ (with $d$ as system dimension) and is related to $\tau_q$ by a Legendre transformation $f(\alpha)=q \alpha -\tau_q$. The number of such points in a lattice scales as $N^{f(\alpha)/d}$. Following from the definition, $f(\alpha)$ is a convex function and satisfies a symmetry  $f(d-2\alpha)=f(\alpha)+d-\alpha$ \cite{evers}. This in turn implies a symmetry in anomalous dimension too: $\Delta_q=\Delta_{1-q}$.

For the delocalized wavefunctions  $f(\alpha)$ is fixed: $f(\alpha)=d$ but its spread increases in cross-over from the delocalized wave limit to the localized one. In case of an ensemble, $f(\alpha, e) =   \lim_{N \to \infty} f(\alpha, e, N) $ can be expressed in terms of the distribution  $P_u(u, e)$ of the local intensity  $u =N \; |\psi|^2$ of a typical eigenfunction $\psi$ \cite{krav}
\begin{eqnarray}
f(\alpha, e, N)  = \frac{d \; \ln(N \; u \; P_u(u, e)) }{\ln N}
\label{fa1}
\end{eqnarray}
where $\alpha=d \left(1- \frac{\ln u}{\ln N} \right)$  with $d$ as the system-dimension and 
$P_u(u, e)={1\over N}  \langle \sum_{k=1}^N \delta(u-N |z_{nk}|^2) \delta(e-e_k) \rangle$.  
For systems with weak multifractality, $f(\alpha)$ is believed to be  approximately parabolic \cite{evers}: $f(\alpha)= d-\frac{1}{4 \epsilon} (d+\epsilon -\alpha)^2 +o(\epsilon^4)$  with $\epsilon \ll 1$. This in turn implies $D_q \approx d-\epsilon \; q$. Note, $d=1$ for a classical ensemble as well as BE.

For a classical ensemble, the eigenfunction are delocalized in the basis-space and $P_u(u)=\int P_u(u,e) \; de$ with $P_u(u)$ as a chi-square distribution \cite{me}: $P_u(u) =  \frac{{\rm e}^{-u/2}}{\sqrt{2\pi u}}$ for XOE and $P_u(u)={\rm e}^{-u}$ for XUE (with X=G, L).
%
%
%
The corresponding $f(\alpha)$ is then 
\begin{eqnarray}
f(\alpha, N) & \approx & 1+ {\beta \over 2} \left(1 - \alpha - {N^{1-\alpha}\over \ln N} \right) + {(\beta-2) \over 2} {\ln 2\pi \over \ln N}
\end{eqnarray}


%
%
To derive $Y$-dependence of $f(\alpha, e, N)$ for a BE, we  first  invert  the relation (\ref{fa1}) which gives $P_u(u,e)=N^{\alpha-2 +f} = {\rm e}^{(\alpha-2+f) \ln N}$. 
As discussed in \cite{pslg}, a variation of the parameter $Y$ gives rise to the diffusion of $P_u(u,e)$ (using the notation $\langle u \rangle_e = {N \over \xi}$):
\begin{eqnarray}
 {\partial P_u \over\partial Y}={2 \; {\mathcal K}_2} \left[{N \over \xi} \; 
   {\partial^2  ( u \; P_u) \over\partial  u^2}  
 +  {\beta \over 2} \; {\partial \over\partial u} \left(u-{N\over \xi} \right) P_u \right] +  L_e P_u 
\label{pu}
 \end{eqnarray}
 where   
\begin{eqnarray}
L_{e} & \equiv & {\partial \over \partial e}\left[
  \beta   a(e)  + {2 \beta \;  N\over E_c} \; e^{\nu} +  {\partial \over \partial e} \; e^{\nu}\right] 
\label{lek}
\end{eqnarray}
with $a(e)=(\frac{2}{\beta})^{\nu} \; e+\frac{\nu}{2}(N-N_a-1)$,  $E_c$ as the Thouless energy and $\nu=0, 1$ for Gaussian and Wishart type Brownian ensembles respectively.

%
%
A substitution of $P_u(u, e)$ as a function of $f(\alpha)$ in eq.(\ref{pu}) leads to the diffusion equation for $f(\alpha)$:
\begin{eqnarray}
   {\partial f_{\alpha}  \over\partial \Lambda_f} \approx  
  {N^{\alpha}\over \xi } \; \left[ 
 {1\over  \ln N } \; \frac{\partial^2 f_{\alpha} }{\partial  \alpha^2} 
 +  \left(\frac{\partial f_{\alpha} }{\partial \alpha} \right)^2  + \frac{\partial f_{\alpha} }{\partial \alpha} \; \left(1+ {\beta \over 2} -{\beta \over 2} {\xi \over N^{\alpha}} \right)\right] +  {\beta \over 2} {N^{\alpha}\over \xi} + T_e f_{\alpha}. \nonumber \\
\label{falp}
 \end{eqnarray}
with 
\begin{eqnarray}
\Lambda_f = {2\; {\mathcal K}_2 \; (Y-Y_0) \over \ln N } 
\label{almf}
\end{eqnarray}
where $T_e$ is the differential operator
\begin{eqnarray}
 T_e f_{\alpha} \equiv   {\ln N \over 2 \; {\mathcal K}_2}  \left( {\beta \; \phi_{\nu} \over \ln N}  +\left(\beta  \; (\phi_{\nu} \; e +\theta_{\nu}) + 2 \; \nu  \right) \; {\partial f_{\alpha}  \over \partial e} + e^{\nu}\; {\partial^2 f_{\alpha} \over\partial e^2}   + \ln N \; e^{\nu} \;  \left({\partial f_{\alpha}  \over \partial e}     \right)^2 \right)
\label{falp1}
 \end{eqnarray}
where $\theta_{\nu}, {\phi}_{\nu}$ depend on the nature of BE: 
$\theta_0= {2 N \over E_c}, \phi_0=1$ for Gaussian BEs, $\theta_1={N-N_a-1 \over 2}, \phi_1= {2\over \beta} +{2 N \over E_c}$ for Wishart BEs.
The appearance of $T_e f$ in eq.(\ref{falp}) clearly indicates an energy-sensitivity of the multifractality spectrum: it is non-stationary along the energy axis. 

A desirable next step would be to solve the above equation but it is technically complicated. 
To gain further insight, we first simplify eq.(\ref{falp}) by a local spectral averaging  which gets rid of the $T_e f$: integrating  eq.(\ref{falp}) over the energy range $e-\Delta e \to e+\Delta e$, while assuming $f$ to be locally stationary over the region, leads to 
\begin{eqnarray}
  {\partial {\overline f}_{\alpha}  \over\partial \Lambda_f} \approx 
  {N^{\alpha}\over \xi} \; \left[ 
 {1\over  \ln N } \; \frac{\partial^2 {\overline f}_{\alpha} }{\partial  \alpha^2} 
 +  \overline{\left(\frac{\partial  f_{\alpha} }{\partial \alpha} \right)^2}  + \frac{\partial {\overline f}_{\alpha} }{\partial \alpha} \; \left(1+ {\beta \over 2} \right)  + {\beta \over 2} \right]  
-{\beta\over 2} \left( \frac{\partial {\overline f}_{\alpha}  }{\partial \alpha}  
 - {\phi_{\nu} \over {\mathcal K}_2}  \right)
\label{falp2}
\end{eqnarray}
where ${\overline f}_{\alpha} = {1\over 2 \Delta e} \; \int_{e-\Delta e}^{e+\Delta e} \; f_{\alpha} \; {\rm d}e$. 
Based on size-dependence of $\xi$, the above equation can further be reduced to a simple form. Noting that $\xi \propto (\langle I_{2} \rangle)^{-1} \propto \; N^{D_2}$ in the spectrum-bulk, with $0 \le D_2 \le 1$),  
%
%
we can approximate, for  $N^{\alpha} \ll \xi $, or equivalently for $\alpha < D_2 \le 1$,  
\begin{eqnarray}
  {\partial {\overline f}_{\alpha}  \over\partial \Lambda_f} \approx 
-{\beta\over 2} \left( \frac{\partial {\overline f}_{\alpha}  }{\partial \alpha}  
 - {\phi_{\nu} \over {\mathcal K}_2}  \right)  .
\label{falp3}
\end{eqnarray}
This indicates a linear $\alpha$ dependence of $f(\alpha)$ for regions $\alpha < D_2$: ${\overline f}_{\alpha} = l_0 + l_1 \; \alpha $ where 
$l_0(\Lambda_f)$ and $l_1(\Lambda_f)$ depend on the initial conditions:
$l_0(\Lambda_f)={\beta \over 2} \; \left({\phi_{\nu} \over {\mathcal K}_2} - l_1 \right) \Lambda_f+l_0(0)$ and $l_1=constant$.

%
%

For regions where $\alpha >> D_2$, the first term  with square bracket of eq.(\ref{falp2}) dominate the 2nd term. This in turn leads to following condition on the possible solution: 
\begin{eqnarray}
 {1\over  \ln N } \; \frac{\partial^2 {\overline f}_{\alpha} }{\partial  \alpha^2} 
 +  \overline{\left(\frac{\partial  f_{\alpha} }{\partial \alpha} \right)^2}  + \frac{\partial {\overline f}_{\alpha} }{\partial \alpha} \; \left(1+ {\beta \over 2} \right)  + {\beta \over 2} = 0.
\label{falp4}
\end{eqnarray}
Thus ${\overline f}_{\alpha}$ now must satisfy  both eq.(\ref{falp3}) as well as eq.(\ref{falp4}) simultaneously;  
one possible solution  in this case seems to be $\overline f_{\alpha} = h_0 + h_1 \alpha$ with 
$h_0=-\frac{\beta}{2}(h_1-{\phi_{\nu} \over {\mathcal K}_2}) \Lambda_f+h_0(0), h_1= -\frac{\beta}{2},-1$.  A linear $\alpha$-dependence of ${\overline f_{\alpha}}$  was indicated also by a previous study \cite{krav} in context of BEs appearing between Poisson to GOE. 



As mentioned above, previous studies of multifractal states have suggested a parabolic solution for $f_{\alpha}$ in weak multifractality regime (with $D_2 \approx 1-2 \epsilon$). Following from eq.(\ref{falp}), such a solution can exist in a  small neighborhood of $\alpha \sim 1 - 2\epsilon + s$  with $s$ given by  the size dependence of $\Lambda_f$: $\Lambda_f = \Lambda_0 \;  N^{-s}$. This can be seen by a  substitution of ${ f}_{\alpha} = v_0 + v_1 \; \alpha + v_2 \; \alpha^2$  in eq.(\ref{falp}) directly (assuming local stationarity) which gives $v_0(\Lambda_f)=\frac{2c}{ln N}\; ln (\frac{1}{v_2(0)}- \lambda_f)+\; \frac{cx\left({v_1(0)}^2 \right)}{1-x\Lambda_f}+ \left(\frac{\beta c}{2}+{\beta \;\phi_{\nu} \over {2\mathcal K}_2}-c{d_0}^2 \right) \lambda_f, v_1(\Lambda_f)=\frac{(v_1(0)+d_0)x}{1-x \lambda_f}+d_0, v_2(\Lambda_f)=\frac{v_2(0)}{1-x \lambda_f}$ where $c=N^{\alpha-D_2}=N^s$, $x=4cv_2(0)$, $d_0=\frac{\beta +2}{4}-\frac{\beta}{4c}$ and $v_k(0)$ with $k=0,1,2$ correspond to initial conditions.

. 



\subsubsection{Diffusion of wavefunction correlations}


%
%
%
%
As intuitively expected, the anomalous scaling behavior of the multifractal states is also reflected by the  overlap of their intensities.  For example, during metal-insulator transition, two wavefunctions say $\Psi(r)$ and $\Psi'(r')$  are known to display following correlation:  $N^2 \langle |\Psi^2(r) \Psi'^2(r') | \rangle \sim \left(|r-r'|\over L_{\omega} \right)^{\Delta_q}$ for $|r-r'| < L_{\omega}$ with  $\Delta_q$ as the anomalous dimension, $L_{\omega} \sim (\rho \omega)^{-1/d}$,  $\omega=|e_i-e_j |$, $\rho$ as the average level density and $d$ as the system dimension \cite{evers}. It is therefore natural to seek the role of the correlations in context of criticality in BEs \cite{evers}.



The two-point intensity correlation $C(e',e'')$  between two eigenstates, say $\Psi_a$ and $\Psi_b$ with eigenvalues $e_a, e_b$ respectively,  for a $N \times N$ matrix $H$, can be defined as  
\begin{eqnarray}
C(e', e'') =\sum_{a,b} \sum_{m=1}^N |\Psi_{ma} |^2 \; |\Psi_{mb}|^2 \; \delta(e'-e_a) \delta(e''-e_b)
\label{corr}
\end{eqnarray}
(with $\Psi_{ma}$ implying $m^{th}$ component of the eigenfunction $\Psi_a$).  
As intuitively expected, its ensemble average  is related to the 2-point spectral correlation $R_2(e',e'')$. This in turn connects   the eigenfunction statistics in the critical regime to that of eigenvalues.
%
%
%
%
As discussed in \cite{pslg} for BEs,  the perturbation by a stationary ensemble leads to an evolution of  $\langle C(e, \omega) \rangle$ from an arbitrary initial condition which depends on both $e, \omega$ (with $e'=e+\omega, e''=e-\omega$) and is non-stationary.  But for the local correlations i.e those for which a variation with respect to $e$  can be ignored, the $Y$-governed evolution  can be approximated as
\begin{eqnarray}
2 \; {\partial  \langle {C} \rangle \over\partial \Lambda_e} &\approx&\left[  {\partial^2     \over \partial r^2}    + \beta  {\partial \over \partial r} \left(2 \eta \; r +{1\over r} \right) -  {(\beta+2)\over 2 \; r^2} + 2 \beta \eta \right] \; \langle {C} \rangle + { \beta \over  4 \; r^2}  \; \langle I_{2(r_0+r)} + I_{2(r_0-r)}\rangle  \; { R}_2(r_0, r)  \nonumber \\
\label{cee3}
\end{eqnarray}
where $\eta= \rm{e}^{- \nu} \; \Delta_e^2 \; \beta_2$ with $\beta_2= \left(\left({2\over \beta}\right)^{\nu} + {\nu  N \over E_c} \right)$,   
$\nu=0, 1$ for Gaussian BE and Wishart BE, respectively, $r_0, r$ are the rescaled energy $e= r_0 \; \Delta_e, \omega = r \; \Delta_e$ with $\Lambda_e$ defined in eq.(\ref{alm1}) and  $I_{2, r}$ is the 2nd inverse participation ratio at energy $r$. 

In the stationary limit $\Lambda_e \to \infty$, it is easy to check that $\langle C \rangle =  R_2 (r_0, r)$  (using the relation $\langle I_{2, r_0} \rangle = {(2+\beta)\over\beta N}$ for the stationary ensembles with ergodic eigenfunctions). An exact solution of the above equation for finite, non-zero $\Lambda_e$ is complicated but, for small-$r$, it  can be obtained by expanding $\langle {C } \rangle$ in Taylor's series around $r=0$. As discussed in \cite{pslg}) the small-$r$ behavior of $\langle C \rangle$ depends on the small-$r$ behavior 
of  $R_2(r)$. For bulk regions where $\langle I_2(r) \rangle$ is almost constant and $R_2(r) \propto r^{\beta}$, one has $\langle C \rangle  \propto r^{\beta}$.  



For criticality considerations, an asymptotic behavior of  $\langle {C} \rangle$ is relevant which can be 
given as $\langle {C} \rangle = r^{-t} \; \sum_{n=0}^{\infty} \; c_n(\Lambda_e) \; r^{-n} $ with coefficients $c_n$ depend on initial conditions and energy-range $r_0$. For  $r_0$ in the  bulk of spectrum,  $ I_{2, r_0+r} = I_{2, r_0-r} \approx I_{2, r_0}$ is almost constant. 
Neglecting  the terms containing $\eta$, due to being $o(1/N)$ smaller as compared to other terms (note $\eta \propto \Delta_e^2$), 
this leads to three possible solutions corresponding to $t=0, 1, 2$:
\begin{eqnarray}
\langle {C} \rangle = {1 \over r^t} \left( c_0(\Lambda_e) + {c_1(\Lambda_e) \over r} + O({1 \over r^2}) \right).
\label{clar}
\end{eqnarray}
where
(i) $c_0(\Lambda_e)=c_0(0)$, $c_1(\Lambda_e)=c_1(0)$ for $t=0$, 
(ii)  $c_0(\Lambda_e)=c_0(0)$, $c_1(\Lambda_e)= \left( c_1(0)  + {\beta\over 4} \; \int_0^{\Lambda_e}  I_2 \; {\rm d}\Lambda_e\right)$ for $t=1$, 
(iii)  $c_0(\Lambda_e)=\left( c_0(0)  + {\beta\over 4} \; \int_0^{\Lambda_e}  I_2 \; {\rm d}\Lambda_e\right)$, $c_1(\Lambda_e)=c_1(0) $ for $t=2$. 
Higher $c_n$ are given by the recursion relation
\begin{eqnarray} 
c_{n+2}(\Lambda_e)  ={\rm e}^{\beta \eta \Lambda_e/2} \left( c_{n+2}(0)  + {\beta\over 4} \; \int_0^{\Lambda_e}  g(\Lambda_e) \; {\rm d}\Lambda_e\right)
\label{cc3}
\end{eqnarray} 
where  $g(\Lambda_e) =
 \left[ 2 (n+t+\beta+1) (n +t) + (\beta -2) \right] \; c_n(\Lambda_e) +   \beta \; I_{2, r_0} \; \delta_{n0} \; \delta_{t0}$.
 where $\delta_{uv}$ is the Kronecker delta function: $\delta_{uv}=1$ or $0$ for $u=v$ and $u\not=v$, respectively.


As discussed above, $\langle C(r_0,r) \rangle \to  R_2(r_0,r)$ for small-$r$. It is therefore appropriate to consider the  measure $K(r) = {\langle C(r) \rangle \over R_2(r)}$ as the criteria for criticality: $K(r) \to 1$ for $r  < 1$ and is  universal  but is system-dependent for  $r  > 1$ (as in this case $R_2 \to 1$ leading to $K(r) \to \langle C(r) \rangle$). For criticality considerations, therefore, the large-$r$ behavior is relevant. 
%
%
For many systems undergoing the localization to delocalization transition of eigenstates, the behavior of $K(r)$ for $r > 1$ is described by the Chalker's scaling \cite{chalk}: $K(r) \sim r^{D_2-1}$ but $K(r) \sim r^{-2}$ for $r$ of the order of spectral band width \cite{krav1}.  But, as clear from the above, the large $r$ behavior of $K(r)$ for a BE depends on the initial conditions as well as location of the spectral scale $e$; here $K\sim {1 \over r^2}$ behavior can occur for $r > 1$ in the bulk spectral regimes   (as here the ensemble averaged inverse participation ratio is almost energy-independent). For BE cases near the edge or intermediate spectral region , $K \sim {c_0 \over r^t} +O(r^{t+1})$ with $t$ determined by the energy-dependence of the inverse participation ratio $I_2$.   

For critical BE cases, $\Lambda_e$ is $N$-independent and some of the higher $c_n$ may become larger than $c_0$. The $K(r)$ behavior in the range $r \sim o(1)$ around $r_0$ is then dominated by 
${1\over r^n}$ term. As an example, we consider the  BE case with Poisson initial condition and in the bulk of spectrum for cases with $I_2 ={N^{-D_2}}$ with $D_2 <1$ and $E_c \sim 1$. As for Poisson limit 
$\langle C \rangle = {1\over N}$ \cite{krav1}, this implies $t=0$, $c_0(0)={1\over N}, c_n(0)=0$ for $n >0$. From the above, we then have $c_0(\Lambda_e)={1\over N} $,
$c_{2} (\Lambda_e)={ \beta \Lambda_e \over 4}  N^{-D_2}$,  $c_{2n}(\Lambda_e)  \sim (\Lambda_e)^n \; N^{-D_2}$ for $n >1$ and $c_{2n+1}(\Lambda_e) =0$ for $n \ge 0$. 
For a size-dependent $\Lambda_e $, say $\Lambda_e \sim  N^{-a}$ such that $D_2+a <1$, 
therefore, the dominant contribution comes from the terms  $r^{-2}$ which leads to  $K(r, \Lambda_e) \sim {1\over r^2}$  for  $r \sim o(1)$.  But for a size-independent $\Lambda_e$,  $c_n$ rapidly increase with $n$ for $n >2$; this  in turn leads to $K(r, \Lambda_e) \sim {1\over r^t}$ with $t$ subjected to the condition  $c_{t+1} < r \; c_t $ and $c_t$ given by eq.(\ref{cc3}).





\section{Critical BE during Poisson $\rightarrow$ GOE transition: numerical analysis}





  The theoretical results in sections II-IV are applicable to  the critical Brownian ensembles
of both Gaussian and Wishart type. For the numerical analysis, however, we focus on a specific Gaussian BE, namely, the one which appears during Poisson to GOE crossover (due to its relevance in context of localization to delocalization transition of the eigenfunctions).



Consider the transition in Gaussian ensembles with an initial state $H=H_0$ described by the ensemble density $\rho_0(H_0) \propto {\rm e}^{-\sum_i H_{0;ii}^2}$.  For a  complete localization of its eigenfunctions in the basis  in which $H_0$ is represented, the initial spectral statistics belongs to the  Poisson universality class.  The perturbation, of strength $\lambda$, by a matrix $V$ taken from a GOE (when represented in the unperturbed basis and of variance $v^2=1$), subjects eigenfunctions to increasingly delocalize as a function of $\lambda$. The ensemble of matrices $H=\sqrt{f}( H_0+ \lambda V)$, with $f=(1+\lambda^2)^{-1}$  then corresponds to the Brownian ensemble during Poisson $\rightarrow$ GOE transition; it is described by the probability density \cite{fgm, ls, alt, shapiro,pich,to}. 
\begin{eqnarray}
\rho(H)  \propto 
{\rm exp}{\left[- {\gamma_b \over 2} \; \sum_{i=1}^{N} H_{ii}^2 -  
2 \gamma_b (1+\mu) \sum_{i,j=1; i < j}^{N} |H_{ij}|^2 \right]} 
\label{be1}
\end{eqnarray}
with $2 (1+\mu)=(\lambda^2 f)^{-1}$ and arbitrary $\gamma_b$; here $H=H_0$ for $\lambda \rightarrow 0$ or 
$\mu \rightarrow \infty$ and $H =V$ for $\lambda \rightarrow \infty$ or 
$\mu \rightarrow 0$. As mentioned in section II, the evolution of matrix elements is described  in terms of the parameter $Y=- {1\over 2} \; \log f$ which in this case becomes $Y \approx {1\over 2 \mu}$. 

%
%
%
%
%

The standard route for the spectral statistical analysis is based on the fluctuations around the  average level density.  In the present case, the  ensemble averaged level density $R_1(e)$, also known as $1^{\rm st}$ order spectral correlation, changes from  a Gaussian to a semi-circular form at the scale of $ N \mu \sim R_1^2$:  $R_1(e) = {N\over \sqrt{\pi}}{\rm e}^{-e^2},  \frac{1+\mu}{\pi} {\sqrt{\frac{2 N}{1+\mu}-e^2}}, N F(e,a) $ for $ (\mu/N) \rightarrow \infty,  0,  a$ respectively \cite{shapiro},
with $a$ as an $N$-independent constant. 
%
Although the exact form of the function $F(e, a)$ is not known, our numerical analysis,  
displayed in figure 1, suggests   a semicircle behavior in the spectral bulk i.e. $F(e) \approx (N b\pi)^{-1} \sqrt{2 b N-e^2}$  with Gaussian tails and $b$ as a constant independent of $N$. (Note the study \cite{shapiro} gives $R_1(e)$ for $H$ as a complex Hermitian matrix  but the numerical evidence given in \cite{psand} and in the present study confirms its  validity also for the real-symmetric $H$.)
Clearly $R_1(e)$ is non-stationary as well as non-ergodic \cite{bg}; as discussed below, this plays a crucial role in compressibility calculation.

As mentioned in section II,  the spectral fluctuations around $R_1(e)$  are governed by the parameter $\Lambda$ \cite{psand}, given by eq.(\ref{alm1}), which in this case becomes, with $Y-Y_0= {1\over 2 \mu}$ and mean level spacing $\Delta_e(e)= R_1(e)^{-1}$,
\begin{eqnarray}
\Lambda_e(e)=  \frac{R_1^2(e)}{ 2 \mu }.
\label{alm}
\end{eqnarray}
For finite $N$, the $\Lambda_e$-variation due to changing $\mu$  at a fixed energy $e$ results in a cross-over of the spectral statistics from Poisson ($\Lambda_e \rightarrow 0$) to GOE ($\Lambda_e \rightarrow \infty$) universality class. 
%
%
%
%
In limit $N \rightarrow \infty$ and for arbitrary $\mu$, $\Lambda_e(e)$ varies abruptly, approaching either $0$ or $\infty$, ruling out possibility of any intermediate statistics. But if $\mu$ takes a value such that the limit $\Lambda^*(e) \equiv \lim_{N \rightarrow \infty} \Lambda_e(e)$ exists, the statistics is then size-independent and  belongs to a new universality class, different from the two end-points and is referred as the critical Brownian ensemble.  As $N$-dependence of $R_1$ also varies with $\mu$, this implies the  existence of two critical points  (instead of one as previously discussed in \cite{alt, shapiro}): 
%


  
\vspace{0.1in}

\noindent {\bf $\mu=c_2 N$}:  as mentioned above, $R_1(e)$ for this case behaves as a semi-circle in the bulk: $R_1(e)= (b \pi)^{-1} \; \sqrt{2 b N- e^2}$.  Although the behaviour near the edge is not known, the numerical analysis, displayed in figure 1 for $c_2=1$, indicates a $\sqrt{N}$-scaling behaviour in all regions: ${1\over \sqrt{N}} \; R_1 \left({e \over \sqrt{N}} \right)$ is $N$-independent. 
Eq.(\ref{alm}) then gives 
\begin{eqnarray}
\Lambda_e(e)=\frac{2 b N-e^2}{2 \pi^2 b^2 N c_2}
\label{alm2}
\end{eqnarray}
 with $b \sim 2$.    
Note, for $c_2=1$, although $\Lambda_e(e)$ is size-independent near  the band-center $e \sim 0$, it is still quite large ($\Lambda \approx \frac{1}{2 \pi^2}$), indicating  the level-statistics to be close to the GOE. An intermediate statistics between Poisson and GOE can however be seen near $e \sim e_0 \sqrt{N}$ for $e_0 \approx 1.7 < b$.

 As mentioned in section IV.1, $\xi \sim N \; \sqrt{Y-Y_0}$ for Gaussian type BEs which  gives, for this case, $\xi \approx \frac{N}{\sqrt{2 \mu}} \sim N^{1/2}$ and $\langle I_2 \rangle \sim \xi^{-1} \sim N^{-1/2}$ in the bulk. 
%
This further implies $\tau_2=D_2 =0.5$ and $\chi=(1-D_2)/2 =0.25$ for the spectrum bulk which is in near agreement with our numerical result (which gives $\tau_2 \approx 0.6$ and $\chi \approx 0.2$, see figure 6(b)). 




\vspace{0.1in}

\noindent $\mu=c_1 N^2$:  as here $\lim_{N \to \infty} {\mu\over N} \to \infty$, $R_1$ now becomes ${N \over \sqrt{\pi}} \; {\rm e}^{-e^2}$. From eq.(\ref{alm}), $\Lambda_e$ is again size-independent: 
\begin{eqnarray}
\Lambda_e(e)={1\over 2 \pi c_1}{\rm e}^{-2e^2}
\label{alm3}
\end{eqnarray}
For $c_1 \sim 1, e \sim 0$, $\Lambda_e \sim {1\over 2 \pi}$ and the statistics lies between Poisson and GOE even for energy ranges  near $e \approx 0$.
 As here $Y-Y_0 \propto N^{-2}$,  this gives $\xi \sim N^0 = O(1)$, $\langle I_2 \rangle \sim N^0$, $D_2 \sim 0$ which suggest a strong multifractal behavior (approaching localization) of the eigenfunctions; (note the latter  rules out the validity of the relation  $D_2 = 1-2 \chi$ in this case). 

The theoretical formulations of the spectral compressibility and multifractal spectrum discussed in previous sections  are based 
on a few approximations at various stages of the derivation. It is therefore  desirable to verify the results by numerical route.
The latter can also give an insight in critical point behavior of some other measures e.g nearest neighbor spacing distribution. 
 The numerical evidence for the criticality for the case $\mu \propto N^2$, with $H$ taken from a real-symmetric ensemble or 
complex-Hermitian ensemble, is discussed and verified in \cite{psand}.
%
The criticality of BE for this case but $H$ taken from a real-quaternion ensemble  was numerically verified 
in \cite{pssymp} (see figure 3 of \cite{pssymp}). In the present work, we pursue a numerical analysis of the
case $\mu \propto N$ only. To understand the non-stationary aspects of critical statistics, we analyze three energy regime i.e . edge, bulk ($e \sim 0$) or at intermediate energies (the region where $R_1(e)$ is half of its maximum value).
%
Although, due to rapid change in $R_1(e)$, the edge results are believed to be error-prone and thus a bit unreliable, but our results  show a systematic trend which encourages us to include them in the figures here. 


\subsection{Critical spectral statistics}

Our theoretical claim about criticality of BE at $\mu=c_2 N$ is based on a $\sqrt{N}$-dependence  of the average level density $R_1$.  Our first step is therefore to numerically confirm its size-dependence.  At this stage, an important question is regarding the ergodicity of the level density for the BE which implies $\rho_{sm}(e)=R_1(e)$, with $\rho_{sm}$  as the spectral averaged level density; $R_1(e)$ can then be used as a substitute for $\rho_{sm}(e)$ for various analytical purposes \cite{bg}. The ergodicity  is confirmed  in a previous study \cite{ss1} (by a  numerical comparison of the ensemble and the spectral averaging of the level density). It is therefore sufficient to analyze the size-dependence of $R_1(e)$.  For this purpose, we consider the ensembles consisting of a  large number of real-symmetric matrices, for many matrix sizes with $c_2=1$; the spectrum for each such ensemble is numerically generated using LAPACK subroutine based on an exact diagonalization approach. As shown in figure 1, $R_1(e)$ is indeed semi-circle in the bulk but 
deviating from it near the edge. Further the $N$-dependence is same for all energy ranges including  edge as well as bulk.

As a next step, we analyze the spectral statistics which requires a careful unfolding of the spectrum. Due to unavailability of the analytical form 
of $R_1(e)$ for all energy ranges,  we apply the local unfolding procedure \cite{gmrr} based on following steps:  the smoothed level density 
$\rho_{sm}$ for each spectrum is first determined by a histogram technique, and then integrated numerically to obtain the unfolded eigenvalues 
$r_n=\int_{-\infty}^{e_N} \rho_{sm} \; {\rm d}e$. The spectrum being non-stationary with energy-sensitive fluctuations (see figures 2,3 of 
\cite{ss1}), it is necessary to  analyze the statistics at different energy-ranges.  For $\Lambda_e$-based comparisons, ideally one should consider an ensemble averaged fluctuation measure at a given energy-point $e$ without any spectral averaging. But in the regions where $\Lambda_e$ varies very
slowly, it is possible to choose an optimized range  $\Delta e$, sufficiently large for good statistics but keeps mixing of different statistics 
at minimum.  We analyze  5${\% }$ of the total eigenvalues taken from a range $\Delta e$, centered at the energy-scale of 
interest i.e. edge, bulk and  intermediate energies.  (As for $\mu=c_2 N \; (c_2=1)$, $\rho_{sm}$ in the bulk is almost constant, the statistics is locally stationary and one can 
take levels within larger energy ranges without mixing the statistics. A rapid variation of $\rho_{sm}$ in the edge however permits one to consider 
the levels  within very small spectral ranges only.  For edge-bulk comparisons, it is preferable to choose the same number of levels for both spectral
regimes). The number of matrices $M$ in the ensemble for each matrix size $N$ is chosen so as  to give approximately $10^5$ eigenvalues and their 
eigenfunctions for the analysis.
%
%





To verify size-independence of the spectral statistics for $\mu \propto N$,  we  consider $P(s)$ and $\Sigma_2(r)$ for the BE with $\mu \propto N$ for many system sizes. For comparison, it is useful to give their behavior in the two stationary limits: 

(i) {\it GOE:} \hspace{0.5in} $P(s)= \frac{\pi}{2} \; s \; {\rm exp} \left(-{\pi} s^2/4 \right), \hspace{0.3in} \Sigma_2(r) =  
\frac{2}{\pi^2} \; \left( {\rm ln} r + C \right)$, with $C \approx 2.18 \label{GOE}$, 

(ii) {\it Poisson:} \hspace{0.22in}  $P(s)=  {\rm exp}(-s), \hspace{0.96in} \Sigma_2(r) = r$. 

It is desirable to compare the BE-numerics with theoretical BE results too but  the exact $P(s)$ behavior  for the BE with matrices of arbitrary size $N$ is not known. It is however easy to derive the $P(s)$ for $N=2$ case \cite{to, psbe}: $P(s, \Lambda) \approx \left({\pi \over 8 \Lambda}\right)^{1/2} \; s \; {\rm e}^{-s^2/16 \Lambda} \; I_0\left({s^2 \over 16 \Lambda}\right)$ with $I_0$ as the modified Bessel function. As $P(s)$ is dominated by the nearest neighbor pairs of eigenvalues, this result is 
a good  approximation also for  $N \times N$ case, especially in small-$s$ and small-$\Lambda$-result \cite{to}. 



 Figures 2, 3 and 4 display the behavior of $P(s)$ and $\Sigma_2(r)/r$ for the BE case $\mu=N$ for many system sizes ranging from $N=500$ to $N=25000$, in three  energy regions.  
With $R_1(e) \propto \sqrt{N}$ for arbitrary $e$ (see fig.1), $\Lambda_e$ (given by eq.(\ref{alm2})) in this case is $N$-independent but its value  varies 
from edge to bulk:  $\Lambda_{e}(e \sim 2.5)  < \Lambda_e(e \sim 1.7 \sqrt{N}) < \Lambda_e(e \sim 0)$.  As a consequence, the statistics is expected  to be critical (i.e intermediate between Poisson and GOE)  but different in the three spectral regimes. This is indeed in agreement with the  behavior of the measures shown in  fig.2, 3, 4. For $e \sim 0$, the statistics is  nearer to  GOE regime (fig.2(c,f), fig.3(c) and fig.4(c,f)) but its deviation from GOE increases for $e \sim e_0 \sqrt{ N}$ case with $e_0 \sim 1.7$ (fig.2(b,e), fig.3(b) and fig.4(b,e)). For $e$ near the edge, the statistics is expected to be  closer to Poisson limit. Although this is confirmed by the tail behavior of $P(s)$ shown in  figure 2(d) and $\Sigma_2(r)/r$ in  3(a) and 4(a,d), the small-$s$ behavior of $P(s)$ is still far from Poisson limit (fig.2(a)). This clearly indicates the dependence of the speed of transition on the spectral ranges: although $\Lambda_e$ is  small in this regime but for spectral ranges $\delta e  < \Lambda_e$,  the transition  to GOE is almost complete. 

The study \cite{shk} suggests that an exponential decaying tail of the $P(s)$ is an indicator for the critical spectral statistics. Fig.2(d,e,f) show a comparison of the tail behavior of $P(s)$ with the curve $P(s) = a \; s \; {\rm exp}(-b s^2 - \kappa s)$ where \textbf{$\kappa \sim  0.70, 0.66, 0.154 $} for levels taken from the edge, intermediate and bulk respectively. (Note, the fit is a close approximation of the  theoretical formulation for $P(s)$ mentioned above for a $2\times 2$ BE). 

The compressibility $\chi$ can be numerically obtained from the large-$r$ limit of $\Sigma_2(r)/r$ curves in figures 3,4; the numerical result is  closer to our theoretical prediction  $\chi=1-4 \sqrt{2} \; \pi \Lambda^*(e)$ (from eq. \ref{cth}). Using $\Lambda^*(e)= {(4-e_0^2) \over 8 \pi^2}$ (from eq.{\ref{alm2}}), we get $\chi=0.11, 0.75$ for the  bulk ($e_0=0$) and intermediate regime ($e_0 \approx 1.7$) respectively. 
(The lack of information about exact $R_1$ in the edge handicaps us from a theoretical prediction for $\Lambda_e$ and therefore $\chi$).
 The small deviations from theory for smaller $N$ can be attributed to the spurious fluctuations due to finite size effects which affects the long-range statistics more severely. The  true fluctuations are expected  to be seen by going to $N \rightarrow \infty$ limit. As can be seen from figure 4, $\Sigma_2(r)/r$ for $N=25000$ are closer to theory than $N=10000$.  Note the bulk-value of $\chi \approx 0.11$ is  expected on the basis of  relation $\chi=(1-D_2)/2$ too (valid for weak multifractal states in the bulk) \cite{chalk}; the latter gives $\chi \approx 0.2$ with our numerically obtained $D_2 \approx 0.6$. For partially localized states, $\chi$ has been suggested to be related to exponential decay of $P(s)$ too \cite{shk}: $\chi \approx {1\over 2 \kappa}$; using $\kappa \approx 0.66$ for intermediate regime $e \sim 1.7 \sqrt{N}$ (given by P(s) fitting mentioned above), this gives $\chi \approx 0.75$ which is again in agreement with our theory.  (Note the range of validity of the relations $\chi \approx {1\over 2 \kappa}$ and $\chi = (1-D_2)/2$ is different; the former is not applicable in near GOE regime and the latter is not valid in strong multifractal regime).   

It must be emphasized  that $\Sigma_2(r)$ results are sensitive to the number of levels  used for the analysis and the ensemble size $M$ even for $N \sim 2.5 \times 10^4$; figure 4 displays the change in behavior for different number of levels taken from a given regime for a given $N$. As the  compressibility calculation is based on  a large $r$ limit of ${\Sigma_2(r) \over r}$, its numerical evaluation for the ensembles of BE type (with rapidly changing level density) can not be reliable.

\subsection{Multifractality analysis of wavefunctions}

Our next step is to investigate the wavefunction statistics based on standard 
measures i.e  inverse participation ratio (IPR),  singularity spectrum and wavefunction correlations at two different energies.
 
In past,  it has been  conjectured that  the distribution  of $I_q$ normalized to its typical value $I_q^{typ} = {\rm exp}\langle \ln I_q \rangle$ has a scale-invariance  at the localization-delocalization transition. This corresponds to  a shape-invariance of $P(\ln I_q)$ with increasing system size $N$, the latter causing only a shift of the distribution along $I_q$ axis \cite{evers}. 
The above conjecture was questioned at first but confirmed later by numerical studies on  Anderson transition for $d >2$ case (with $d$ as dimension) and critical power law random banded matrix (PRBM).  To check its validity in case of the critical BEs,  
we numerically analyze  the eigenstates for the case $\mu=N$. 
To overcome finite size effects, one has to consider averages over different realizations of disorder as well as  
a narrow energy range. As these fluctuations in bulk are analyzed in detail in \cite{krav}, here we confine ourselves to
intermediate regime only. For this purpose, we consider the eigenstates in a narrow energy range 5${\% }$ around intermediate energy for each matrix of the ensemble with $\mu=N$, consisting of $M$ matrices, with $M=8000, 6000, 5000, 3000, 2500, 1500$ for $N=500, 750, 1000, 1500, 2000, 3000$ respectively.
%
%
Figure 5(a) shows the distribution $P(\ln I_2)$ for the critical BE with $\mu= N$; the scale invariance of the distribution  
is clearly indicated from the figure. As indicated by previous studies \cite{evers},  the $I_q$-distribution is expected to show a power-law tail  at the transition :
$P(I_q/I_q^{typ}) \propto (I_q/I_q^{typ})^{-1-x_q}$ for $I_q \gg I_q^{typ}$; the behavior is confirmed in figure 5(d) for $q=2$ with $x_{q=2}\gg 1$; (our numerics gives $x_2 \sim 100$ however a more detailed analysis is needed due to huge errors possible in tail of the distribution). 
%
Furthermore the change in peak-position  of $P(\ln I_2)$ with changing system size confirms a power-law dependence of  $\langle I_2 \rangle$ on 
system size $N$, governed by a continuous set of exponents: $\langle I_2 \rangle \sim N^{-\tau_2^{typ}}$ where $\tau_2^{typ}=\tau_2$ for $x_2 >1$.

%
As mentioned in section IV, the  multifractal behavior of eigenfunction is described by  a continuous set of scaling exponents $\tau_q$  \cite{evers}. The latter can be computed by standard box-size scaling approach. This is based on first dividing the system of $L^d$ 
basis states into $N_l =(L/l)^d$ boxes ($d$ is the dimension of the system and for our case, d=1) and computing the box-probability $\mu_k$ of $\psi$ in the $i^{th}$ box : $\mu_k(l) =\sum_n |\psi_n |^2$; 
here $\sum_n$ is over basis-states within the $k^{th}$ box. This gives the  scaling exponent $\tau_q$ for the typical average of
$I_q(l)=\sum_{k=1}^{N_l} \mu_k^q(l)$: 
\begin{eqnarray}
\tau^{typ}(q)  = \frac{\langle \ln I_q(\lambda) \rangle }{\ln \lambda} 
\label{tau}
\end{eqnarray}
where $\langle . \rangle$ is the average over many wavefunction at the criticality.  For numerical calculation of $\tau_q^{typ}$, one usually considers 
the  limit $\lambda \equiv l/L \rightarrow 0$ which can be achieved either by making $L \rightarrow \infty$ or $l \rightarrow 0$.  We 
choose $\lambda=0.1$ and carry out $\tau_q$ analysis for many $N$ values, each considered for an ensemble size $M=20$; (a large ensemble size $M$ is not required for their analysis). 
%
%
For case $\mu=N$ and $q >0$, the slope of  $\tau_q $ vs $q$ curve turns out to be $1/2$ which gives $D_q \approx 0.5$  (see figure 5(b)) which agrees well with our theoretical prediction (see eq.(\ref{tau1})).  This is also confirmed in figure 5(c) displaying $N$-dependence of $\overline{\langle I_2 \rangle} $ which is well-fitted by the expression $\overline{\langle I_2 \rangle}(e_0 \sqrt{N},N) \approx 8 e_0 $. Rewriting in terms of $e$, this implies $\overline{\langle I_2 \rangle}(e,N) \approx 8 {e \over \sqrt{N}}$ and therefore  reconfirms $D_2 \approx 0.5$.  Note, our result for $D_q$ is in contrast with the study \cite{krav}  which theoretically predicts $D_2 \approx 2-\gamma$ for $\mu \propto N^{\gamma}$  but  numerical verifies the result only for the cases $\gamma \not= 1$).




Next we numerically analyze the singularity spectrum using box-approach in 
which $f(\alpha)$ and $\alpha$  are defined as follows \cite{romer}: 
$\alpha_q^{typ} =\lim_{\lambda \rightarrow 0} \frac{1}{\ln \lambda} \langle {\frac{1}{I_q(\lambda)} \; \sum_{k=1}^{N_{\lambda}} {\mu_{k}}^q(\lambda) \; \ln \mu_k(\lambda) }\rangle $ and 
\begin{eqnarray}
f(\alpha_q^{typ}) =\lim_{\lambda \rightarrow 0} \frac{1}{\ln \lambda} \left[ q \langle \frac{1}{I_q(\lambda)} \; \sum_{k=1}^{N_{\lambda}} {\mu_{k}}^q(\lambda) \; \ln \mu_k(\lambda) \rangle -\langle ln \; I_q(\lambda) \rangle \right]
\label{ftyp}
\end{eqnarray}
with superscript $''typ''$ on a variable implying its typical value. It is believed that the typical spectra is equal to the average spectra (i.e. $\tau_q^{typ}= \tau_q$ and $f^{typ}(\alpha)=f(\alpha)$) in the regime $q_{-} < q < q_{+}$ \cite{evers}. Here $q_{\pm}$ correspond to the values of $q$ such that $f(\alpha_q)=0$; the corresponding value of $\alpha_q$ are  referred as $\alpha_{\pm}$, respectively.   Our numerics of $f(\alpha)$ is confined within this regime. 
 As  displayed in figure 5(f)  for six system sizes, $f(\alpha)$ behavior for the case $\mu=c_2 N$ is intermediate between  the localized and delocalized limit. Also clear from the figure, $\alpha$ is contained in  the interval $(0, 2)$ and $f(\alpha)$ satisfies the symmetry relation $f(2-\alpha)=f(\alpha)+1-\alpha$.
The symmetry  $\Delta_q= \Delta_{1-q}$ in the spectrum of  $\Delta_q$ can also be seen from the figure 5(e). 
Our analysis gives $\alpha_0 = 1.3 > d, \alpha_1= 0.74, f(\alpha_0)= d= 1, f(\alpha_1)= \alpha_1$. Above results are consistent with expected multifractal characteristics of the critical eigenstates \cite{romer, evers}. 

The non-stationarity of the spectral statistics and existence of non-zero correlations between eigenfunctions and eigenvalues suggest the multifractality measures to be sensitive to chosen energy-regime. This is also indicated by our theoretical analysis (see eqs.(\ref{falp1}, \ref{falp2}) however a local spectral averaging almost hides the energy-dependence of $f(\alpha)$.  The main reason for this could be attributed to stronger sensitivity of the measures $\tau_q$, $I_2^{typ}$  to $N$-dependence. Figure 6 compares the ensemble averaged $\tau_q$, $I_2^{typ}$ as well as singularity spectrum for three different energy ranges; although the energy dependence of $I_2^{typ}$ is clear from fig.(b) but nearly same behavior of $\tau_q$, $f(\alpha)$ indicates an almost insensitivity of these measures to the energy-scale. This  is in contrast to spectral measures $P(s)$ and $\chi$ where the non-stationary effects are more pronounced.  


To reveal the non-stationarity effects on  the eigenfunction fluctuation, it is therefore necessary to consider a  measure in which energy-scales play an important role. As discussed in section IV.3,  the 2-point wavefunction correlation is one such measure.  Here we numerically analyze $\langle C(e, \omega) \rangle$, given by eq.(\ref{corr}), for $20 \%$ energy levels chosen in bulk ($e \sim 0$) as well as in the intermediate-edge spectral regime.  As discussed in section IV.3, the behavior of $\langle C\rangle$ is expected to change near $\omega \sim E_c$, with its curvature changing sign. Using the definition $E_c \sim \Delta_e \; N^{D_2}$, with $\Delta_e \propto N^{-1/2}$ and $D_2 =0.5$, one has $E_c \sim 1$.  As displayed in figure 7, the curvature of ${\langle C \rangle}$-curve indeed changes sign  near $\omega \sim 1$, with $\langle C \rangle$  increasing for $\omega \le 1$ and then undergoes a power law decay for  $\omega > 1$. The decay however is faster than $1/r^2$ in both the regimes.  As $\Lambda_e$ in this case is size-independent, this is in agreement with  theoretical prediction (see end of section IV.3). 
The figure also displays different decay rates in the two regimes which is  
expected due to different spectral rate of variation of $\overline{\langle I_2 \rangle}$ in the bulk and intermediate; as can be seen from fig.8(b), 
$\overline{\langle I_2 \rangle}$ is almost constant in the bulk but increases rapidly around $e \sim N^{0.6}$. This confirms the  sensitivity of  $\langle C(e, \omega) \rangle$ to the  energy-regime of interest.

 

In the end, we compare our results for various critical measures with those in study \cite{krav}. For an ensemble density described by eq.(\ref{be1}) with $\mu \propto N^{\gamma}$, the theoretical analysis of  \cite{krav} predicts (i) $D_q = 2-\gamma$ for $q>1/2$, (ii) $f(\alpha)= {\alpha\over 2} +1 -{\gamma\over 2}$ for $\alpha_{min} < \alpha < \gamma$; here $\alpha_{min}$ depends on $\gamma$: $\alpha_{min} =0, 2-\gamma, \gamma$ for $\gamma >2$ and $2 > \gamma > 1$ and $\gamma \le  1$ respectively, (iii) $K(\omega) \sim {1\over \omega^2}$ for $\omega > E_{c}$ for all $\gamma$.
Our theoretical analysis gives following results for the same ensemble:  (i) $D_q = (2-\gamma)/2$ for spectrum bulk for $q>1/2$, (ii) a linear $f(\alpha)$ for $\alpha < D_2$ and $\alpha >> D_2$ but possibility of a parabolic behavior near $\alpha \sim 1$, (iii)  $K(\omega) \sim {1\over \omega^2}$ for $\omega > E_{c}$ only in bulk  and for $1 < \gamma < 2$ (the latter corresponds to a size-dependent $\Lambda_e$ with  $N^{1-\gamma} < \Lambda_e \propto N^{2-\gamma}$).  $\Lambda_e$ being size-independent for $\gamma=1, 2$, the large $\omega$-decay of $K(\omega)$ can be faster than ${1\over \omega^2}$. Our theoretical predictions are corroborated by the numerical analysis of case $\gamma=1$. ( Note the study \cite{krav} presents $K(\omega)$-numerics for $\gamma \not= 1 , 2$ only).  The deviation of our $D_2$-result from \cite{krav} may be due to 
their choice of a fixed size-dependence of the mean-level spacing ($ \propto N^{-1}$) for all $\gamma$ while we have used $\Delta_e \propto N^{-\gamma/2}$; the latter  result is  derived  in \cite{shapiro}) and is confirmed by our numerics too (see fig.1).

\section{connection with other ensembles}




A Gaussian Brownian ensemble is a special case of a multi-parametric Gaussian ensemble. 
As indicated by the studies \cite{ps-all,psijmp, psand}, the eigenvalue distributions
of a wide range of ensembles with single well potential e.g  those with a multi-parametric 
Gaussian measure and independent matrix elements,  
appear as a non-equilibrium stages of a Brownian type diffusion process \cite{ps-all}.
Here the eigenvalues evolve with respect to a single parameter, say $Y$, 
which is a function of the distribution parameters of the ensemble. The parameter is  related to the complexity of the system 
represented by the ensemble and can therefore be termed as the spectral "complexity" parameter. 
The solution of the diffusion equation for a given value of the complexity parameter gives the
distribution of the eigenvalues, and thereby their correlations, for the 
corresponding system.  As the local spectral fluctuations are defined on the scale of 
local mean level spacing, their diffusion is governed by a competition between $Y-Y_0$ and local mean level 
spacing. Consequently the evolution parameter $\Lambda_e$ for the local spectral statistics is again given by eq.(\ref{alm1}) but with a 
more generic definition of $Y$; (note  so far the complexity parameter formulation has been analyzed in detail only in context of  
Gaussian ensembles although the studies \cite{ps-all, psnher} indicate its validity for more generic cases). A single 
parameter formulation is also possible for the eigenfunction fluctuations but, contrary to spectral case, the parameter is not 
same for all of them \cite{pswf, ps-all, pslg}.

The implications of the complexity parametric formulation are significant: as the system dependence enters through a single parameter in a fluctuation measure, its behavior for different systems with same value of the complexity parameter (although may be consisting of different combinations of the system parameters) will be analogous 
(valid for same global constraints; see \cite{ps-all} for details).
An important point worth emphasizing here is the following: although the unfolding (rescaling by local spectral density)  of the eigenvalues removes their dependence on the local spectral scale, the latter is still contained in $\Lambda_e$.  The spectral dependence of $\Lambda_e$ varies from system to system. Thus two systems in general may have same spectral statistics at a given spectrum-point but the analogy need not extend for a spectral range of sufficient width. It could however happen in case the two systems have same local rate of change of $\Lambda_e$ along the spectrum which usually requires a similar behavior for the local spectral density. The analogy implied by the complexity parameter formulation is therefore strictly valid only in case of the ensemble averaging. It can however be extended to include spectral averaging within the range in which  the local density is almost stationary.

The Anderson ensemble (AE) consisting of Anderson Hamiltonians, the power law random banded matrix (PRBM) ensemble and the Brownian ensemble appearing during Poisson $\to$ GOE transition belong to  same global symmetry class (time-reversal symmetry preserved). 
Based on the complexity parameter formulation, therefore, the critical point statistics of an AE or PRBME can be mapped to that of the Poisson $\to$ GOE Brownian ensemble.  The validity of the mapping was indeed confirmed by a number of numerical studies  \cite{psand, pssymp}. 
As discussed in \cite{psand, pswf, pssymp}, the critical BE analog of a critical AE is unique; similar to an AE, the level-statistics of the BE shows a scaling behavior too.  
The study \cite{krav} however claims that the critical point behavior  for an Anderson ensemble and a PRBM ensemble differ from that of a Rosenzweig-Porter ensemble (same as the Brownian ensemble between Poisson $\to$ GOE cross-over). For example, the study shows that 
the correlation $C(\omega)$ between two wavefunctions, at energies $e$ and $e+\omega$ decays as $\omega^{-\mu}$ for $\omega \gg E_{th}$, with $\mu=2$ for Rosenzweig-Porter ensemble and $\mu=D_2-1$ for Anderson Hamiltonian and PRBM ensemble.  Here $E_{th} \sim N^{-z}$ is the Thouless energy (same as $E_c$ used in context of BEs), with $z=1$ for AE and PRBME and $z <1$ for the BE.  These results are however based on the assumption of local stationarity of the spectral density around which the fluctuations are measured. The seeming contradiction of the results between \cite{krav} and \cite{psand} originates in the range of validity of the assumption. As indicated by previous studies, the ensemble averaged bulk spectral density of an Anderson ensemble is almost similar in behavior as that of  a PRBM ensemble  but is different from that of the Poisson $\to$ GOE Brownian ensemble.  In the latter case, it varies more rapidly along the spectrum (see section V); the spectral range $r$ of local stationarity in case of  the BE is therefore much smaller than the AE and PRBME and the measures (e.g. compressibility)  which are based on large $r$-limit considerations may not be appropriate for the comparison. Indeed the complexity parameter based formulation permits a comparison of the measures for each spectral point and is therefore more suitable for a comparative analysis of cases with different spectral-densities.

\section{Conclusion}

Based on a non-perturbative diffusion route, we find that the criticality of the fluctuation measures for the  BEs is sensitive to both spectral scale as well as the perturbation strength.   Our theoretical results are applicable for both Gaussian as well as Wishart BEs of the Hermitian matrices, with or without time-reversal symmetry and appearing during transition from an arbitrary initial condition to stationary ensembles. The results are confirmed by a numerical analysis of the BEs appearing during Poisson to GOE transition. The relevance of our BE-results is expected to be wide-ranging.   For example,  BEs are  connected to the ensembles of column constrained matrices and latter has application in many areas discussed in \cite{ss1}.  Further, using the complexity parameter based mapping 
of the fluctuation measures of a BE to a multi-parametric Gaussian ensemble \cite{ps-all}, the results derived  here are useful for the latter too.

An important outcome of our analysis is to reveal a new criteria for the criticality of the random matrix ensembles i.e  the spectral complexity parameter.  The latter has been shown to govern the evolution of all spectral fluctuation measures for a multi-parametric ensemble including BEs \cite{ps-all}; the search for criticality therefore need not depend on a specific measure e.g. compressibility. 
Using the complexity parameter, it is  easier to find the number of critical points too: the spectral statistics has a critical point at a fixed energy if the size-dependence of the perturbation strength $Y$ is same as that of the square of the mean level spacing. The appearance of two critical points in case of the BE between Poisson and GOE (i.e the Rosenzweig-Porter ensemble)  can therefore be attributed to the variation of the level density from a Gaussian to semi-circle form.  This also predicts the existence of two critical points in a Wishart Brownian ensemble which appears during Poisson to WOE transition;this follows because their level density  changes from exponential decay to the $\sqrt{a-e}$ form (with $a$ as a constant, see discussion below eq.(21) of \cite{pslg}). 
The existence of two critical points was recently reported in context of other complex systems too e.g. many body localization as well as random graphs \cite{krav1}. 


The complexity parameter has an another  advantage over previous measures for criticality which were often based on the assumption of the local ergodicity. As the search for the criticality  originated in context of disordered systems, usually with large flat regions in the bulk level density, the local ergodicity 
considerations were easily satisfied. In general however this is not the case e.g. for systems with rapidly changing level densities. The measures based on the ensemble averaging only, or those based on averaging over very small spectral ranges are more appropriate choices to seek  critical point in such cases.

The present work  deals with the BEs taken from  Hermitian matrix space.
An understanding of critical BEs lying between the pairs of stationary ensemble subjected to other global constraints e.g. non-Hermiticity (.e.g. circular ensembles), chirality, column constraints still remains  an open question.

\vfill\eject


\oddsidemargin=-80pt
\begin{figure}
\centering
\vspace*{-40mm}
\includegraphics[width=1.2\textwidth,height=1.8\textwidth]{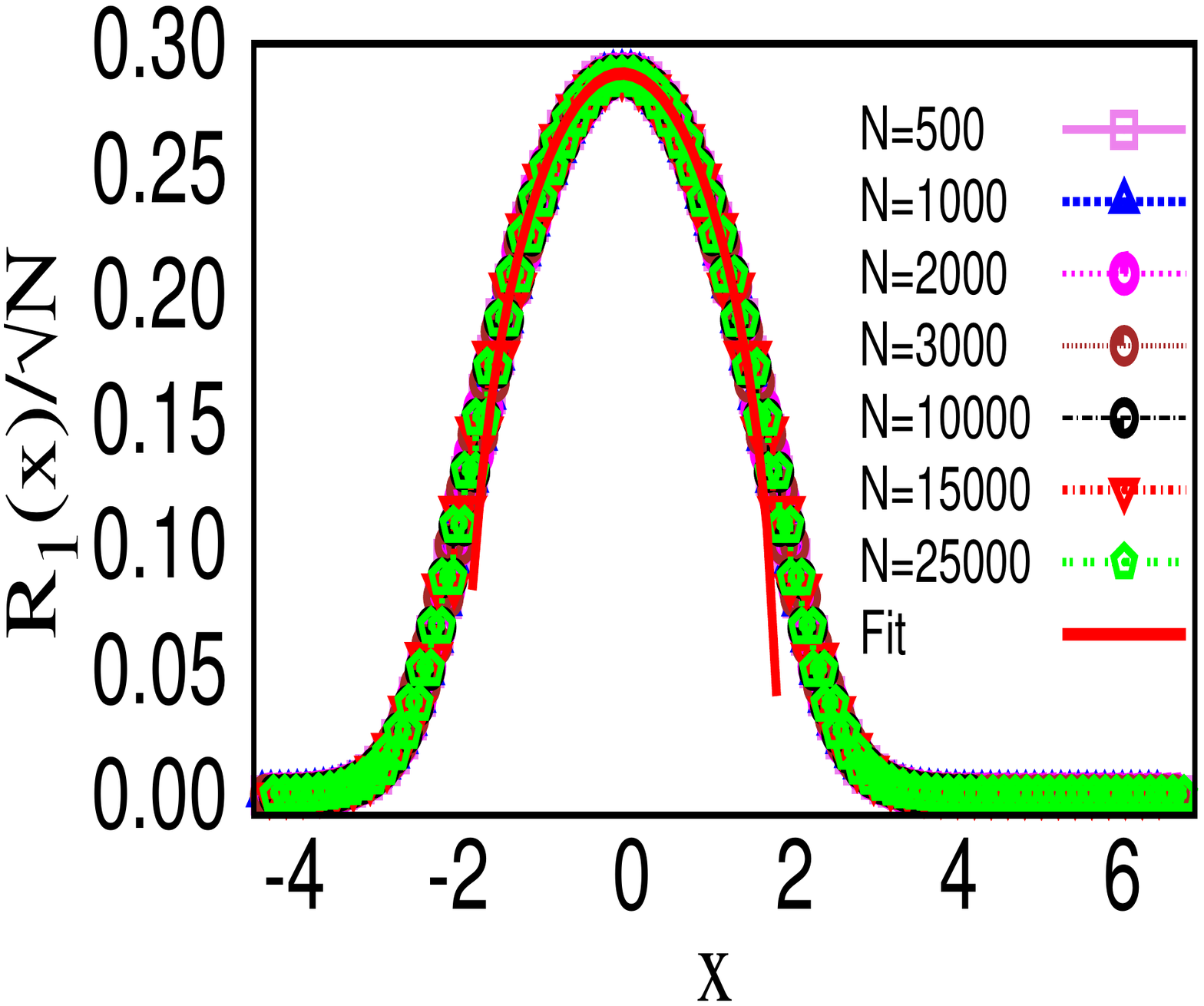} 
\vspace*{-80mm}
\caption{ 
{\bf  Ensemble averaged level density $R_1(x)$}: Behavior of the Brownian ensemble (BE) eq.(\ref{be1}) with $\mu=N$ for 
many system sizes $N$ where $x=e/\sqrt{N}$; here $R_1(x)$ for different $N$ is scaled by $\sqrt{N}$. 
The solid line corresponds to the fit- $R_1(e) ={1\over b\pi } \; \sqrt{2bN-e^2}$ with $b \approx 2$, confirming the semicircle behaviour at bulk. The behavior near the edge is deviating from semicircle fit but collapse of $R_1(x)$ for different $N$ on the same curve indicates same N-dependence for all energy ranges: $R_1(e)=\sqrt{N} f(e/\sqrt{N})$ . A comparison of $R_1(e)$ with 
spectral level density $\rho_{sm}(e)$ is given 
in \cite{ss1}.
}
\label{fig1}
\end{figure}

\oddsidemargin=60pt
\begin{figure}
\centering
\vspace*{-20mm}
\includegraphics[width=1.2\textwidth, height=1.05\textwidth]{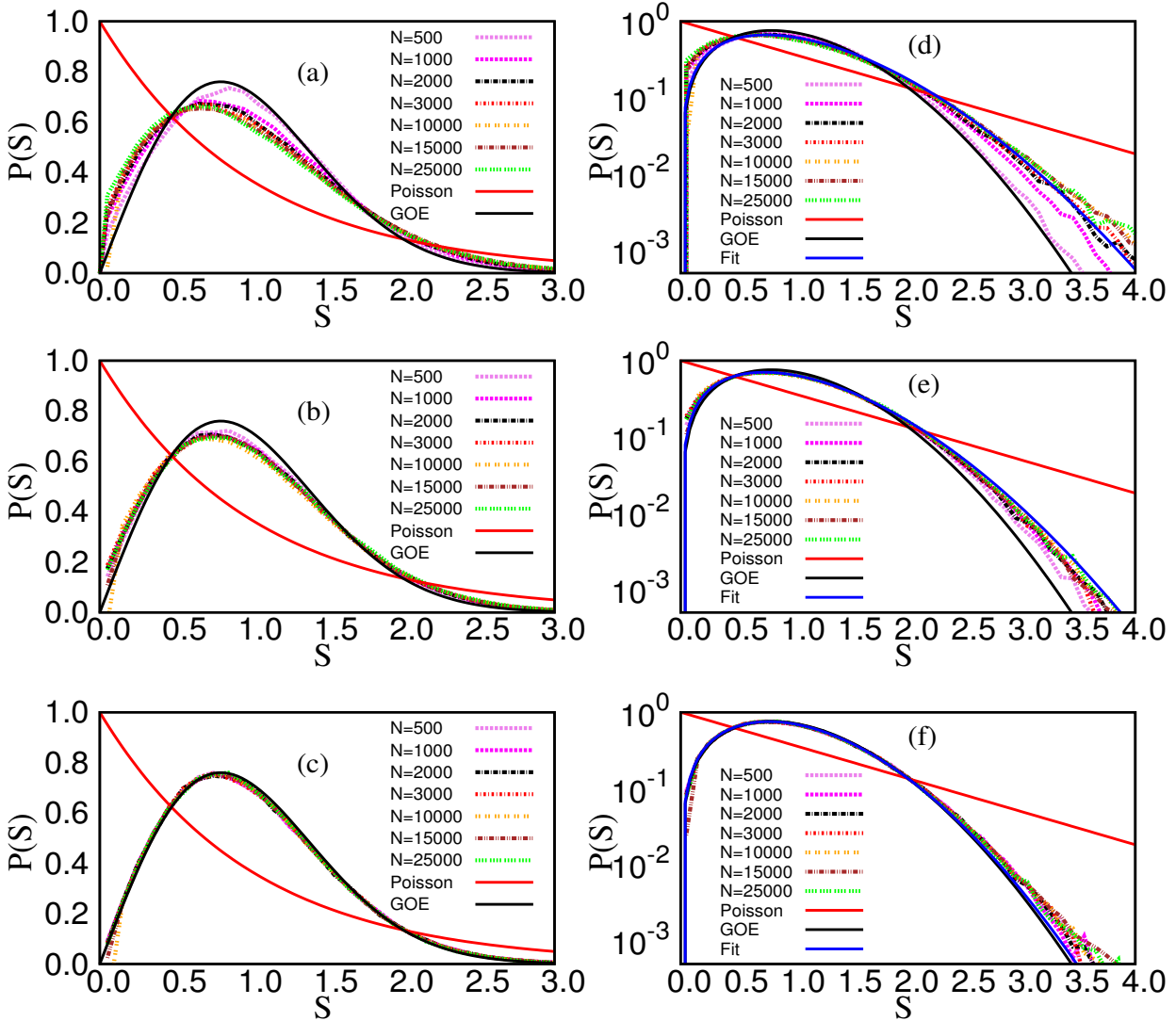} 
\vspace*{-18 mm}
\caption{
{\bf Non-stationarity of $P(s)$ :}
Nearest neighbor spacing distribution for the ensemble density eq.(\ref{be1}) with $\mu=N$ for many system sizes $N$ in three energy ranges: (a) and (d)- edge (neighbourhood of minimum $R_1(e)$),  
(b) and (e)- intermediate (the neighbourhood where $R_1(e)$ is half of its maximum value), (c) and (f)- bulk (neighbourhood of maximum $R_1(e)$). Sensitivity of $P(S)$ to the energy can be seen from the small 's' behavior (fig. (a), (b), (c)) and large 's' behavior (fig. (d), (e), (f)). As clear from fig.(a) and (d), deviation of $P(s)$ from GOE increases as $N$ increases.
 The behavior in the bulk is close to GOE limit but the one in intermediate regime is different from both Poisson and GOE limit (the difference is more clear in fig.(e) although it can also be seen in fig.(b) near $S \sim 1$); 
as $\Lambda_{bulk} > \Lambda_{intermediate} > \Lambda_{edge}$, the above shift of statistics from GOE  is in agreement with theoretical prediction.  As expected for the critical 
statistics, $P(s)$ in (b) approaches an  invariant form with increasing system size $N$. The parts (d), (e), (f) also compare the tail behavior with the fit-
$[a \; s \; {\rm exp}(-b s^2 - \kappa s)]$ with $a=1.9, b=0.42, \kappa= 0.70$ for edge, $a=2.01, b=0.47, \kappa=0.66 $ for intermediate regime, $a=1.7, b=0.73, \kappa= 0.154 $ for bulk. 
}
\label{fig2}
\end{figure}


\oddsidemargin=-10pt
\begin{figure}
\vspace*{-26mm}
\hspace*{-32mm}
\vspace*{-25mm}
\includegraphics[width=1.6\textwidth, height=1.3\textwidth]{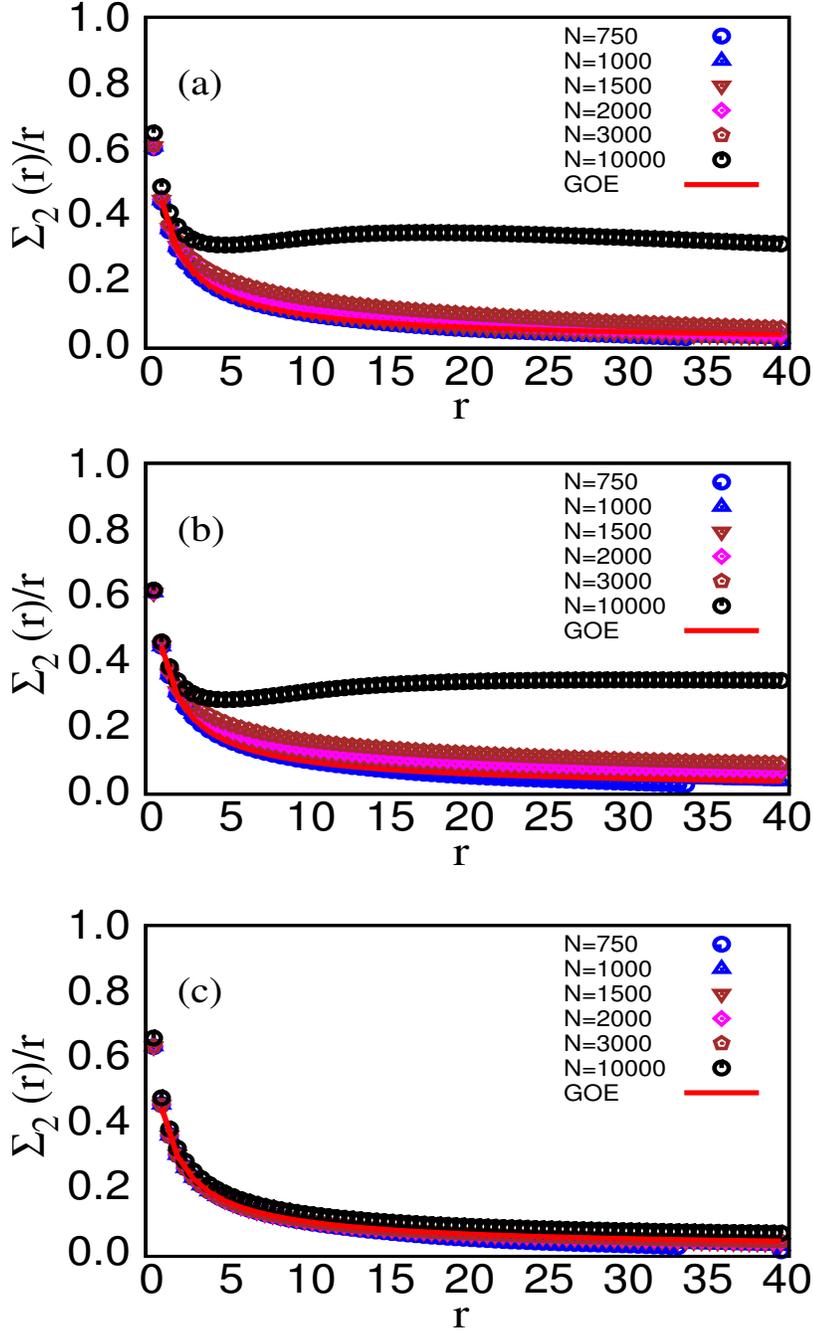} 
\vspace*{3 mm}
\caption{ {\bf Non-stationarity of compressibility $\Sigma_2(r)/r$}: Variance of number of levels in a distance of $r$ mean level spacings for the BE, eq.(\ref{be1}) with $\mu=N$ for many system sizes  in three energy ranges :
 (a) edge, (b) intermediate,  (c) bulk. The solid line in (a, b, c) corresponds to the theoretical prediction for GOE mentioned  in section (\ref{GOE}). 
As indicated by the parts (a) and (b), the critical behavior of $\chi$ (i.e. $0<\chi<1$) is not evident for small N cases but appears only in large $N$ limit; (note however an upward shift of the curves, although very small, can be seen even for small $N$). This is caused by the spurious fluctuations due to finite size effects, expected to be more pronounced in the large $r$-limit. This is analyzed in more detail in figure 4.}
\label{fig3}
\end{figure}

\begin{figure}
\vspace*{-26mm}
\hspace*{-12mm}
\vspace*{-25mm}
\includegraphics[width=1.4\textwidth, height=1.1\textwidth]{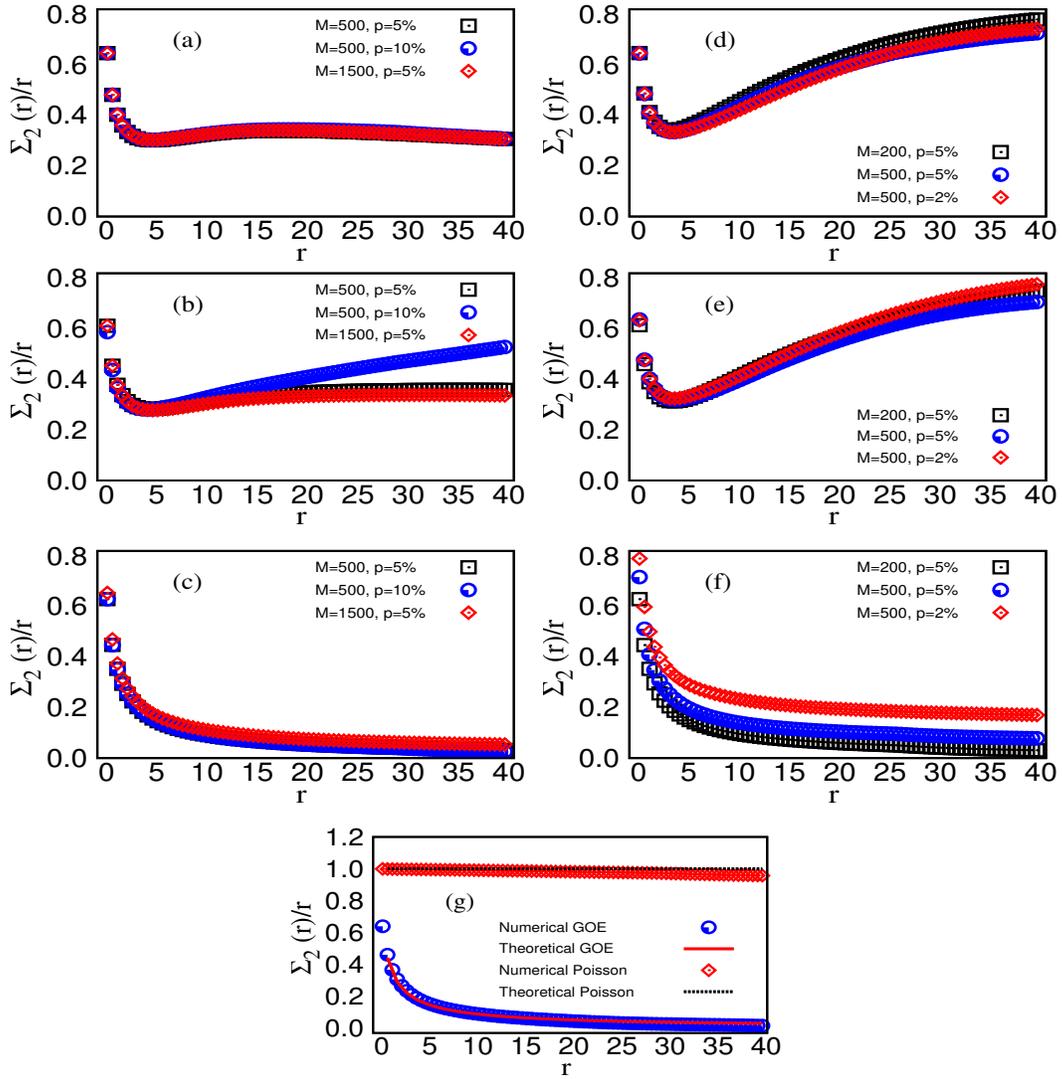} 
\vspace*{-8 mm}
\caption{ {\bf  Finite size effect of the compressibility $\Sigma_2(r)/r$ }:
The sensitivity of the number variance $\Sigma_2(r)$ to size $N$ in a given energy regime  is evident from figure 3. 
To probe it further, here we again consider the behavior for large $N$, namely, for $N=10000$ (figs.(a,b,c)) and $N=25000$ (figs.(d,e,f)) in three different energy regime (edge- fig.(a) and (d), intermediate- fig.(b) and (d), bulk- fig.(c) and (f)); the symbol $"M"$ here refers to the ensemble size (number of matrices taken for one particular $N$) and the symbol $"p"$ refers to the number of levels used for the numerics from the energy regime under consideration. As evident from the figures, the large-$r$ behavior for $N=25000$ approaches to a fractional compressibility ($\approx$ $0.75$ and $0.1$, as expected from theoretical prediction (eqn(\ref{cth}, \ref{alm2})) in the intermediate and bulk regime, respectively). The behavior is however sensitive to '$p$' variation for a fixed '$M$' suggesting the non-stationarity of $\Sigma_2(r)$. As a consequence, it is not easy to implement the large-$r$ limit necessary for the compressibility calculation. 
To validate the efficiency of our numerical code, a comparison of the numerically simulated result for GOE and Poisson ensemble with theory, are shown in part (g).
}

     
%
\label{fig3a}
\end{figure}


\oddsidemargin=-10pt
\begin{figure}
\vspace*{-26mm}
\hspace*{-12mm}
\vspace*{-25mm}
\includegraphics[width=1.2\textwidth, height=1.06\textwidth]{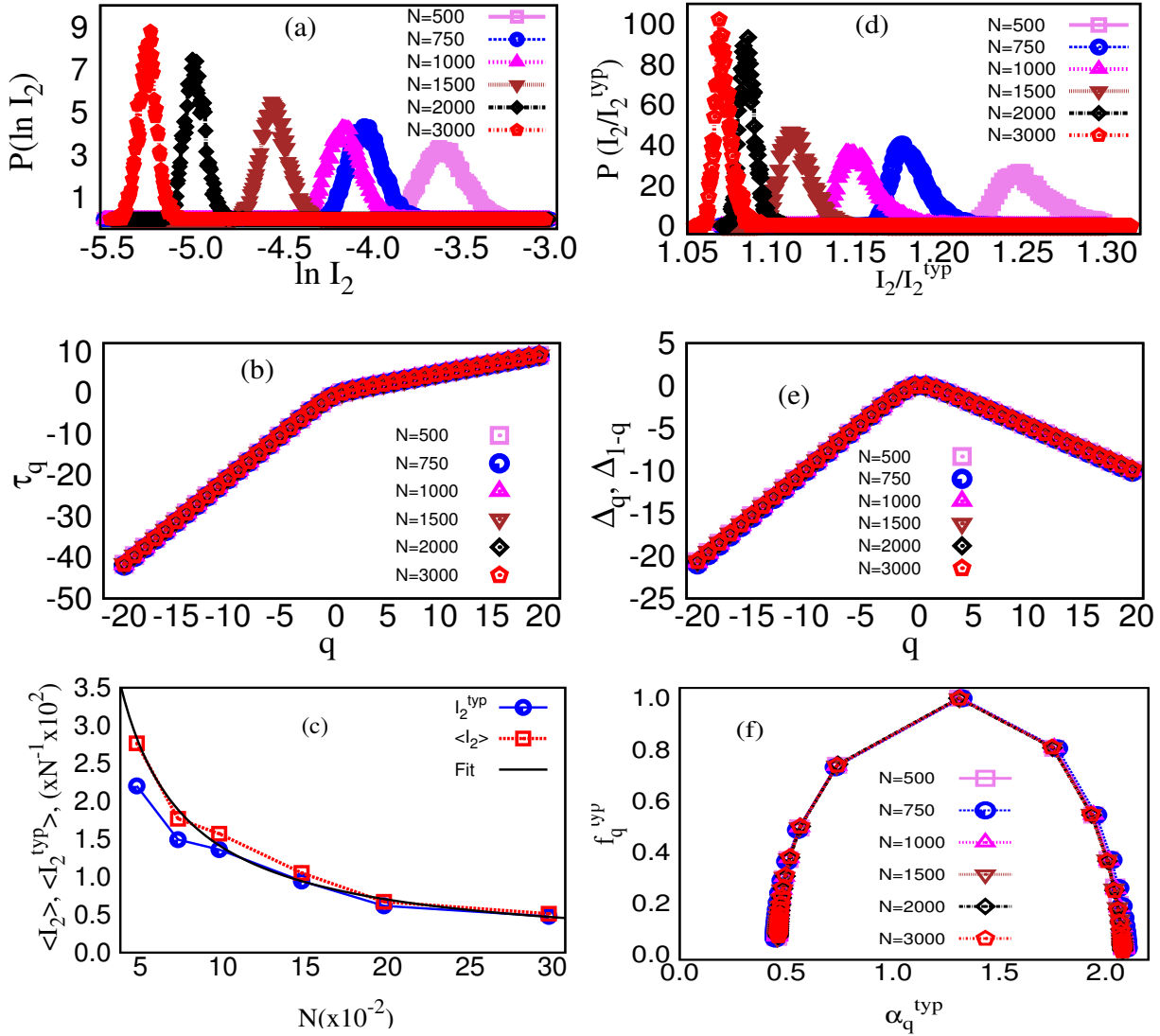} 
\vspace*{6 mm}
\caption{ 
{\bf  Multifractality of eigenfunctions at intermediate regime}: 
The figures display the distribution of IPR $I_2$ (spectral averaged locally) as well as multifractality spectrum for BE eq.(\ref{be1}), with $\mu=N$, for many system sizes, at the intermediate energy regime: (a) $P({\rm ln}I_2)$- distribution shifts along ${\rm ln} I_2$ axis preserving their form as $N$ increases, (b) $\tau_q$ - as clear from the display, the straight line for $q>0$ has a  slope  ${{\rm d} \tau_q \over {\rm d}q} \approx {1\over 2}$  which agrees well with our theoretical prediction eq.(\ref{tau1}) (with $\gamma=1$), 
%
(c)  $ {\langle {I}_2 \rangle}(e_0\sqrt{N}, N)$ and ${\langle{I_2}^{typ} \rangle}(e_0\sqrt{N}, N) $: (for clarity of presentation, here  the rescaled variables ${\langle {I}_2 \rangle \over 100 N}$ and ${\langle{I_2}^{typ}\rangle \over 100 N}$ are displayed with respect to rescaled size ${N \over 100}$). The $\langle {I}_2 \rangle$ curve fits well with ${14.17 \over N}$ which gives $D_2 \approx 0.5$ reconfirming our theoretical prediction (see discussion below eq.(\ref{tau}) for clarification), (d) $P(I_2/I_2^{typ})$ - here the fit $f(I_2)=\left(\frac{I_2}{I_2^{typ}}\right)^{-1-x_2}$ at $I_2\gg I^{typ}_2$  gives $x_2 \gg 1$ (our numerics give $x_2\approx 100$), which in turn implies  $I_2^{typ} = \langle I_2 \rangle$, (e) Anomalous dimension $\Delta_q$ - a symmetry around $q=0$ is evident from the figure (see section IV.2) which also implies the 
symmetry of the singularity spectrum, (f) $f^{typ}(\alpha_q)$ - as suggested on theoretical grounds,  $f^{typ}(\alpha_q)$ (eq.(\ref{ftyp})) seems to approach a linear behavior in the region $\alpha < D_2 \approx 0.5$ and $\alpha > 1.5$ alongwith a parabolic behavior near $\alpha \sim 1$. The theory however predicts a narrowing parabolic regime as $N$ increases. 
}
\label{fig5}
\end{figure}


\oddsidemargin=-10pt
\begin{figure}
\vspace*{-26mm}
\hspace*{-45mm}
\vspace*{-25mm}
\includegraphics[width=1.6\textwidth, height=1.3\textwidth]{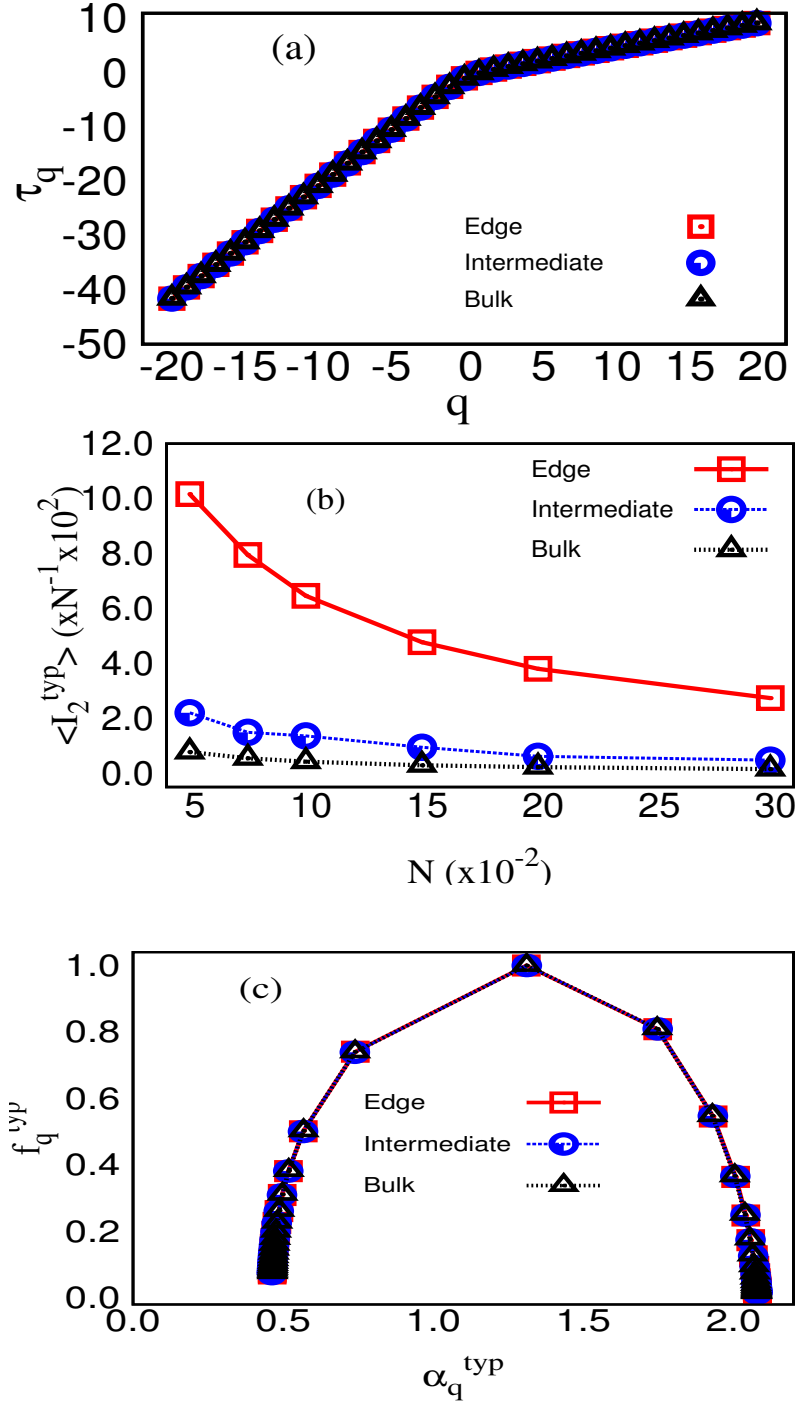} 
\vspace*{20 mm}
\caption{ 
{\bf  Sensitivity of the multifractality to an energy regime}: 
The figures display the multifractality spectrum for BE eq.(\ref{be1}), with $\mu=N$ at three energy regimes. Although the energy dependence of $\langle {I_2}^{typ} \rangle$ is clear from (fig.(b)) but nearly same behavior of $\tau^{typ}(q)$  in fig.(a) (eq.(\ref{tau})) as well as $f_q(\alpha)$ behavior in fig.(c) (both for $N=3000$) for three energy ranges indicates a very weak  sensitivity to energy-range of these measure which is further suppressed due to local spectral averaging.
}
\label{fig5a}
\end{figure}

\oddsidemargin=-10pt
\begin{figure}
\vspace*{-30mm}
\hspace*{-30mm}
\vspace*{-26mm}
\includegraphics[width=1.3\textwidth, height=1.2\textwidth]{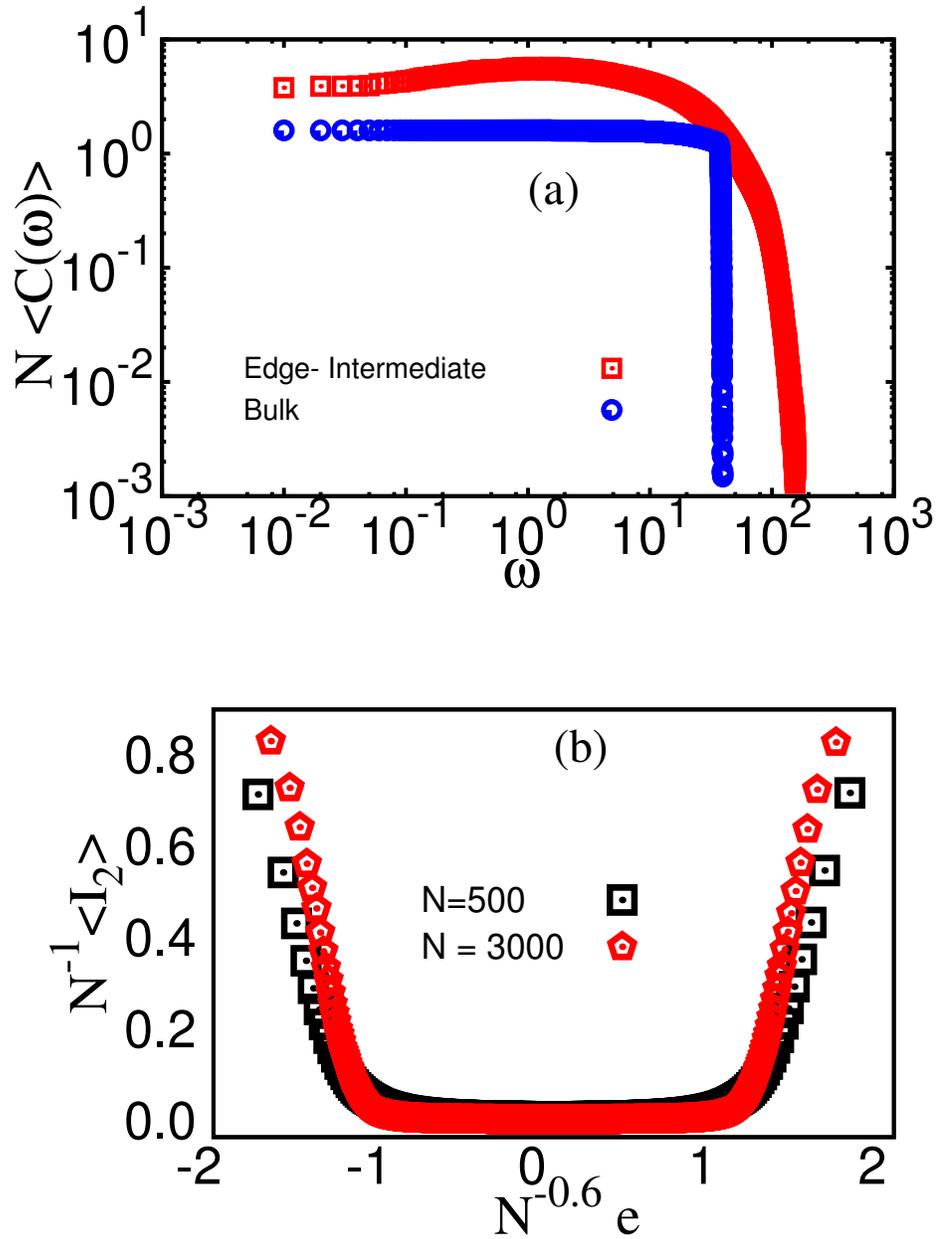} 
\vspace*{8 mm}
\caption{ 
{\bf Non-stationarity  of 2-point intensity correlation}: 
The figures displays the $\langle C(\omega) \rangle$ (eq. (\ref{corr}), for $N=3000$) and $\langle I_2 \rangle$ (for $N=500, 3000$) for BE eq.(\ref{be1}), with $\mu=N$. In fig.(a), the numerics is based on 20${\%}$ levels in the energy range of interest . This leaves us only with two energy ranges for the analysis: "edge-intermediate" (as intermediate regime almost overlaps with edge) and "bulk". The function $N\langle C(\omega) \rangle \propto \omega$ for $\omega < 1$ and undergoes a power law decay for $\omega > 1$, however decay is faster than $1/\omega^2$ as predicted by theoretical calculation in section $(IV.3)$.  
As evident from the fig.(a), the decay rates are different in the two regimes   
which is expected due to non-stationarity of $\overline{\langle I_2 \rangle}$. As discussed in section $(IV.3)$, the energy-dependence of $\langle C \rangle$ comes from $I_2$ which varies rapidly for energy-ranges away from bulk.  This is verified  in fig.(b) which shows an almost constant $\overline{\langle I_2 \rangle}$ in the bulk but a rapid increase around $e \sim N^{0.6}$; (note the figure shows the plot of $N^{-1} \; \overline{\langle I_2 \rangle}$ with respect to rescaled  $e \rightarrow e/N^{0.6}$). This confirms the  sensitivity of  $\langle C(e, \omega) \rangle$ to the  energy-regime of interest.
%
%
}
\label{fig6}
\end{figure}

\end{document}